\documentclass[final,authoryear,12pt,5p,times]{elsarticle}

\usepackage{bm}
\newcommand{\bmath}[1]{\bm{#1}}

\newcommand{\mathbmss}[1]{\bm{\mathsf{#1}}}
\newcommand{\var}{\mathop{\text{Var}}\nolimits}
\newcommand{\cov}{\mathop{\text{Cov}}\nolimits}

\newcommand{\expect}{\mathop{\mathbb{E}}}
\newcommand{\disp}{\mathop{\mathbb{D}}}
\newcommand{\As}{\mathop{\mathbb{A}\text{s}}}
\newcommand{\Ex}{\mathop{\mathbb{E}\text{x}}}
\newcommand{\const}{\mathop{\text{const}}\nolimits}
\newcommand{\erfc}{\mathop{\text{erfc}}\nolimits}
\newcommand{\FAP}{{\text{FAP}}}
\newcommand{\Dw}{\mathop{\text{Dw}}}

 \usepackage{graphicx}

\usepackage{amssymb}
\usepackage{amsmath}

\journal{Astronomy and Computing}

\begin{document}
\sloppy

\begin{frontmatter}


%

\title{Statistical detection of patterns in unidimensional distributions by continuous
wavelet transforms}


\author{Roman V. Baluev}
\address{Central Astronomical Observatory at Pulkovo of the Russian Academy of Sciences,
Pulkovskoje sh. 65/1, Saint Petersburg 196140, Russia}
\address{Saint Petersburg State University, Faculty of Mathematics and Mechanics, Universitetskij
pr. 28, Petrodvorets, Saint Petersburg 198504, Russia}
 \ead{r.baluev@spbu.ru}

\begin{abstract}
Objective detection of specific patterns in statistical distributions, like groupings or
gaps or abrupt transitions between different subsets, is a task with a rich range of
applications in astronomy: Milky Way stellar population analysis, investigations of the
exoplanets diversity, Solar System minor bodies statistics, extragalactic studies, etc. We
adapt the powerful technique of the wavelet transforms to this generalized task, making a
strong emphasis on the assessment of the patterns detection significance. Among other
things, our method also involves optimal minimum-noise wavelets and minimum-noise
reconstruction of the distribution density function. Based on this development, we
construct a self-closed algorithmic pipeline aimed to process statistical samples. It is
currently applicable to single-dimensional distributions only, but it is flexible enough to
undergo further generalizations and development.
\end{abstract}


\end{frontmatter}



\section{Introduction}
Nowdays, the wavelet analysis technique is frequently used in various fields of astronomy.
It proved a powerful tool in the time-series analysis, in particular to trace the time
evolution of quasi-periodic variations \citep{Foster96c,Vit-wav}. So far, the time-series
analysis remains the major application domain of the wavelet transform, and most of the
wavelet methodology and results are tied to this field. However, there are other branches
where this technique appeared promising, in particular in the analysis of statistical
distributions. In multiple astronomical applications we deal with statistical samples and
distributions of various objects. For example, the last $20$ years brought up a rich
diversity of exoplanetary systems, and investigating their distributions gives a tremendous
amount of information about the process of planet formation, their migration and dynamical
evolution \citep{CummingStat}. Other possible applications include the analysis of Milky
Way stellar population that becomes more important with the emerging GAIA data
\citep{Gaia17}, and the statistical analysis of minor bodies distributions in Solar System.

We emphasize that we formulate our goal here as `patterns detection' rather than `density
estimation'. The latter would literally mean to estimate the probability density function
(p.d.f.) of a sample, but instead we aim to detect easily-interpretable structures and
shapes in this p.d.f., like e.g. clusters of objects, or paucities, or quick gradients.
Such inhomogeneties often carry hidden knowledge about physical processes that objects of
the sample underwent. In such a way, our task becomes related to data mining techniques and
cluster analysis.

First attempts to apply the wavelet transform technique to reveal clumps in stellar
distributions date back to 1990s \citep{Chereul98,Skuljan99}, and \citet{Romeo03,Romeo04}
suggested the use of wavelets for denoising results of $N$-body simulations. Nowdays,
wavelet transforms are quite routinely used to analyse CMB data from WMAP
\citep{McEwen04,McEwen17}. The very idea of `wavelets for statistics' is not novel too
\citep{Fadda98,ABS00}.

When applied to these tasks, wavelets allow to objectivize the terms like the `detail' or
`structural pattern' in a distribution, and easily formalize the task of `patterns
detection'. However, there are several crucial issues in this technique that either remain
unresolved or solutions available in the literature look unsatisfactory and sometimes even
flawed. In particular, the following matters raise questions.
\begin{enumerate}
\item Applying discrete wavelet transforms (DWTs) in this task seems unnatural. A major
argument in favour of the DWT in 1990s might be to improve the computing performance.
Nowdays, cumputing capabilities do not limit the practical use of continuous wavelet
transforms (CWTs). Besides, the CWT mathematics is easier in many aspects.
\item Most if not all authors perform preliminary binning of the sample, or another kind of
smoothing, before they apply a wavelet transform. It is an unnecessary and possibly even
harmful step. Small-scale structures are averaged out by the binning, and wavelets cannot
`see' them after that.
\item There is a big issue with correct determination of the statistical significance at
the noise thresholding stage (discussed below).
\item It was not verified, whether the classic and typically used wavelets are indeed
suitable in this task. Perhaps a systematic search is necessary, among wavelets of
different shape and utilizing some objective criteria of optimality.
\end{enumerate}

A crucial problem of any statistical analysis is how to justify the statistical signifiance
of the results. Determination of the statistical significance was always recognized as an
important issue in this task. Unfortunately, approaches developed for the wavelet analysis
of time series are not applicable for distribution analysis. Nonetheless, several schemes
are available in the literature of how to do significance testing of the wavelet
coefficients derived from a statistical sample \citep[e.g.][]{Skuljan99,Fadda98}. But all
these works share one common flaw: they define the significance of an \emph{individual}
wavelet coefficient, but test in turn \emph{multiple} coefficients at once. In practice it
leads to a dramatic increase of the false detections rate above the predicted level.

Consider that we perform $N_t$ independent significance tests on the same sample, and every
individual test is tuned to have a small enough false alarms probability (or p-value)
$\beta$. The total number of false alarms is then $\sim \beta N_t$, and it may become
impredictably large because of large $N_t$. Let $N_d$ be the number of significant
(`detected') wavelet coefficients that passed the test. Then the fraction of false alarms
\emph{among the detected coefficients} is $\beta_{\rm rel}\sim \beta N_t/N_d$. Usually
$\beta_{\rm rel} \gg \beta$, so the relative fraction of false alarms becomes much larger
than the requested `false alarm probability', paradoxically compromising the latter term.
In other words, the false alarm probability is in fact misapplied in this task. In practice
it may easily appear that \emph{the majority} of the wavelet coefficients that formally
passed their individual significance tests, appear in turn just noisy fluctuations.

In applications, the attention is paid to every detected detail of the distribution. Each
false-detected wavelet coefficients trails a false `pattern' in the recovered distribution.
We guess that a researcher would expect that \emph{all} structures that were claimed
significant by the analysis algorithm, are significant indeed. So our intention is to
narrow the `false detection' term from `an individual wavelet coefficient was wrongly
claimed significant' to a more stringent `at least one of many wavelet coefficients was
wrongly claimed significant'. This triggers an effect generally similar to the one known as
the `bandwidth penaly' in the periodogram analysis of time series
\citep{HorneBal86,SchwCzerny98b,Baluev08a}.

In addition to what said above, only Monte Carlo simulations can currently be used to
calculate the necessary p-values when testing the CWT significance. But numerical
simulations are obviously inefficient, because they are very CPU-expensive and lack the
generality. Some basic initial work on this problem was made in \citep{Baluev05-51Peg}. In
this paper we treat analytically the above-mentioned issues, and present the entire
analysis pipeline.

\section{Overview of the paper}
\label{sec_overview}
In the literature there is a deficit of research dedicated to the task stated above, and
there is a diversity of lesser sub-problems yet to be solved. Although many useful partial
results are available, the very formalism of this task is still under construction, so
there is no complete and self-consistent theory that we could use here `as is'. Below we
consider the following issues:
\begin{enumerate}
\item adaptation of the CWT technique to statistical samples and distributions
(Sect.~\ref{sec_wavs});
\item characterization of the noise that appears in the wavelet transform and construction
of the signal detection criterion (Sect.~\ref{sec_noise});
\item control of the non-Gaussian noise in the CWT that appears due to the small-number
statistics and limits applicability of the entire technique (Sect.~\ref{sec_noise});
\item search of optimal wavelets that improve the efficiency of the analysis
(Sect.~\ref{sec_opt});
\item optimal reconstruction of the distribution function itself from its CWT after noise
thresholding (Sect.~\ref{sec_rec});
\item numerical simulations and tests aimed to verify our theoretic results and
constructions, demonstrate the main issues of the technique, and determine limits of its
practical applicability (Sect.~\ref{sec_simul}).
\end{enumerate}
Finally, in Sect.~\ref{sec_alg} we provide a brief summary of our wavelet analysis
algorithm.

\section{Wavelet transforms}
\label{sec_wavs}
\subsection{Basic definitions and formulae}
We adopt the classic definition of the CWT from \citep{GrMorlet84} with only a minor
modification in scaling:
\begin{equation}
Y(a,b) = \int\limits_{-\infty}^{+\infty} f(x) \psi\left(\frac{x-b}{a}\right) dx.
\label{wavdef}
\end{equation}
Here, $f(x)$ is an input function of the CWT. In this paper, it is meant to be a p.d.f. The
kernel $\psi(t)$ is meant to be a \emph{wavelet}. The latter term does not have a stable
and strict definition, but at least $\psi$ must be well localized together with its Forier
transform $\hat\psi$. Classic definitions also contain normalization factors
in~(\ref{wavdef}), typically $1/\sqrt a$, which we discard here.

The integral transform~(\ref{wavdef}) is similar to a convolution, but contains two
parameters: the scale $a$ and the shift $b$. Contrary to the usual convolution, the CWT is
easily invertible. Multiple inversion formulae are available, in particular based on
\citep{Liu15} we can write:
\begin{align}
f(x) &= \frac{1}{C_{\psi\gamma}} \int\limits_{-\infty}^{+\infty}\int\limits_{-\infty}^{+\infty} Y(a,b) \gamma\left(\frac{x-b}{a}\right) \frac{da db}{|a|^3}, \nonumber\\
C_{\psi\gamma} &= \int\limits_{-\infty}^{+\infty} \hat\gamma(\omega)\hat\psi^*(\omega) \frac{d\omega}{|\omega|},
\label{wavinv}
\end{align}
Here, the choice of the reconstruction kernel $\gamma(t)$ is rather arbitrary: it is mainly
restricted by the mutual admissibility condition $0<|C_{\psi\gamma}|<+\infty$. One of the
most famous inversion formulae \citep{GrMorlet84,Vit-wav} contains $\gamma=\psi$, and it is
similar to the original CWT~(\ref{wavdef}). It requires that the wavelet must satisfy the
classic admissibility condition $0<C_{\psi\psi}<+\infty$, implying in particular that
$\hat\psi(0)=0$, and hence $\psi(t)$ must integrate to zero. This special case can be
viewed as an orthogonal projection in the Hilbert space of $Y$, while other $\gamma$
correspond to \emph{oblique} projections.

The generalized inversion formulae~(\ref{wavinv}) can be verified by applying the Fourier
transform to it.

An alternative definition of the CWT can be written down as
\begin{equation}
\Upsilon(\kappa,s) = \int\limits_{-\infty}^{+\infty} f(x) \psi(\kappa x+s) dx,
\label{wavdefkc}
\end{equation}
where $\kappa=1/a$ is a wavenumber-like parameter, while $s=-b/a$ is a phase-like
parameter. In terms of $\kappa$ and $s$, the inversion formula~(\ref{wavinv}) attains the
following shape:
\begin{equation}
f(x) = \frac{1}{C_{\psi\gamma}} \int\limits_{-\infty}^{+\infty}\int\limits_{-\infty}^{+\infty} \Upsilon(\kappa,s) \gamma(\kappa x+s) d\kappa ds.
\label{wavinvkc}
\end{equation}
These alternative definitions differ from the classic ones only in the parametrization of
their arguments, i.e. $\Upsilon(\kappa,s)=Y(1/\kappa,-s/\kappa)$.

We need to discuss practical subtleties of the inversion formulae~(\ref{wavinv}). Their
direct use might seem problematic in practice, because the integration with respect to the
scale parameters $a$ or $\kappa$ is unlimited.\footnote{The infinite integration with
respect to $b$ or $s$ is typically not a problem, since the integrands
in~(\ref{wavinv},\ref{wavinvkc}) tend to zero quickly enough for $b\to\infty$ or
$s\to\infty$.} However, the two types of parametrization infer different infinite ranges:
either $a\to\infty$ (large scales) or $\kappa\to\infty$ (small scales). Which parameters
offer better numerical efficiency, depends on the limiting behaviour of the integrand in
the relevant tail.

The parameters $k$ and $s$ appear more practical in this concern: $\Upsilon(\kappa,s)$ is
always smooth near $\kappa=0$, while $Y(a,b)$ is bad-behaved for $a\to 0$. Therefore, in
practical computations we always represent the CWT in terms of the $(\kappa,s)$ arguments.
However, in the mathematical formulae following below we mainly stick with the traditional
$(a,b)$-notation.

Concerning the integration of the small-scale tail $\kappa\to\infty$, in practice we always
have to limit $|\kappa|$ in~(\ref{wavinvkc}) by some $\kappa_{\rm max}$. This infers that
all the scales smaller than $a_{\rm min}=1/\kappa_{\rm max}$ are eliminated, triggering an
effect of smoothing in the resulting $f(x)$. In other words, the reconstruction of $f(x)$
on the basis of $\Upsilon$ or $Y$ is necessarily approximate. But from what goes below it
follows, that the uncertainties (or noise) in $\Upsilon(\kappa,s)$ reside mainly at the
small scales as well. Loosing some small-scale structure is an internal property of the
noise thresholding procedure constructed below. But this is a natural issue inherited from
an ill-posed nature of our original task: to reconstruct the p.d.f. from a finite sample.
It is not a problem of the inversion formulae~(\ref{wavinv},\ref{wavinvkc}) themselves.

\subsection{Informal justification of the method}
Our goal is to detect specific structural patterns in the sample, such that can be used to
refine our knowledge of the physics behind the sampled objects. Such patterns may include
e.g. clusters or gaps, or sharp gradients. Contrary to the wavelet analysis of time series,
we usually do not expect to have any periodic patterns in the distribution, so the
technique presented below are targeted to reveal aperiodic structures in general.

The potential use of the CWT in this task can be demonstrated as follows. Let us choose
some smoothing kernel $\varphi(t)$ (not a wavelet yet), and apply it to a derivative
$f^{(k)}(x)$, instead of $f(x)$ itself. Then use integration by parts $k$ times to obtain
the following equality:
\begin{equation}
\int\limits_{-\infty}^{+\infty} f^{(k)}(x) \varphi\left(\frac{x-b}{a}\right) dx = \frac{(-1)^k}{a^k} \int\limits_{-\infty}^{+\infty} f(x) \varphi^{(k)}\left(\frac{x-b}{a}\right) dx.
\label{dersp}
\end{equation}
We assume here that $\varphi(t)$ tends to zero for $t\to \pm\infty$, along with its
derivatives, so the relevant boundary terms do not appear in~(\ref{dersp}).

We can see that the right-hand side of~(\ref{dersp}) involves a derivative kernel
$\psi_k(t)=(-1)^k\varphi^{(k)}(t)$. Now assume we aim to reveal local maxima or minima or
zones of fast changes in the p.d.f. $f(x)$. This can be done by considering the derivatives
$f'(x)$ and $f''(x)$. The first derivative carries information about slopes in $f(x)$, and
highlights sudden jumps. The second derivative carries information about deviation of
$f(x)$ from its local linear slope, highlighting groupings ($f''<0$, concave upward $f$) or
gaps ($f''>0$, convex downward $f$). The formula~(\ref{dersp}) provides the necessary link
between these derivatives and the CWT via the wavelets $\psi_{1,2}$.

The function $\varphi$ should desirably have a simple bell-like shape with non-vanishing
integral, so it cannot be a wavelet. We will call $\varphi$ the generating function for
$\psi_k$.

A classic example is given by Hermitian wavelets, which satisfy our construction, and are
derived from the Gaussian:
\begin{equation}
\varphi(t) = e^{-\frac{t^2}{2}} \implies
\psi_1(t) = t e^{-\frac{t^2}{2}}, \,
\psi_2(t) = (t^2-1) e^{-\frac{t^2}{2}}.
\label{wHerm}
\end{equation}
These kernels are sometimes called the WAVE and MHAT (`Mexican Hat') wavelets, and they
infer a Gaussian smoothing on the derivatives $f'(x)$ and $f''(x)$.

\subsection{Estimating the wavelet transform from a sample}
Assume that we aim to analyse the distribution of a random quantity $x$, and this $x$ has
the p.d.f. of $f(x)$. We will refer to it as to `the $x$-distribution'. The
CWT~(\ref{wavdef}) can be viewed as the mathematical expectation of another random quantity
$y$:
\begin{equation}
y=\psi\left(\frac{x-b}{a}\right), \qquad Y(a,b) = \expect y.
\label{ydef}
\end{equation}
We will call its distribution `the $y$-distribution'. It depends on the underlying
$x$-distribution, on the adopted wavelet shape, and also on the parameters $a,b$.

From observations we have a random sample of $N$ values $x_1,x_2,\ldots,x_N$, drawn from
the $x$-distribution, and we need to estimate the CWT based on this sample. Extending the
definition from \citep[sect.~3.2]{ABS00} to the continuous case, we may adopt the sample
mean estimate
\begin{equation}
\widetilde Y(a,b) = \frac{1}{N} \sum_{i=1}^N y_i = \frac{1}{N} \sum_{i=1}^N \psi\left(\frac{x_i-b}{a}\right).
\label{wavest}
\end{equation}
This can be used as a practical formula that allows one to derive the estimation of
$Y(a,b)$ from the sample $\{x_i\}$. We will call~(\ref{wavest}) the sample wavelet
transform (SWT), by analogy with mean/sample mean, variance/sample variance, etc.

For simplicity, let us introduce a shorthand $\langle *\rangle$ for the sample-averaging
operation:
\begin{equation}
\langle \phi \rangle := \frac{1}{N} \sum_{i=1}^N \phi_i \quad \text{or} \quad \langle \phi(x) \rangle := \frac{1}{N} \sum_{i=1}^N \phi(x_i).
\end{equation}
Then~(\ref{wavest}) turns to just $\widetilde Y = \langle y \rangle$.

Obviously, (\ref{wavest}) is an unbiased estimate of~(\ref{wavdef}), i.e. $\expect
\widetilde Y = Y$ for any $a,b$. Moreover, if $f$ and $\psi$ are such that the central
limit theorem is satisfied, $\widetilde Y$ is asymptotically Gaussian for $N\to\infty$. Its
variance can be expressed as
\begin{align}
D(a,b) &:= \disp \widetilde Y = \frac{\disp y}{N} = \nonumber\\
&= \frac{1}{N} \left[ \int\limits_{-\infty}^{+\infty} f(x) \psi^2\left(\frac{x-b}{a}\right) dx - Y^2 \right].
\label{wavvar}
\end{align}
One can see that this function may vary dramatically over the $(a,b)$ plane. This means
that the noise contribution in the estimate $\widetilde Y$ is non-uniform. In other words,
$\widetilde Y$ becomes heteroscedastic, rendering this estimate practically unsuitable
\citep[sect.~5.8]{Jansen-NRWT}. It would be much more useful to deal with a constant noise
level, so we need to normalize $\widetilde Y$ in some way to equibalance the noise over the
$(a,b)$ plane. For this goal, we may use the classic unbiased variance estimate
\begin{align}
\widetilde D(a,b) &= \frac{1}{N-1} \left[ \frac{1}{N} \sum_{i=1}^N \psi^2\left(\frac{x_i-b}{a}\right) - \widetilde Y^2(a,b) \right] = \nonumber\\
 &= \frac{\langle y^2 \rangle - \langle y\rangle^2}{N-1}.
\label{wavvarest}
\end{align}
This is basically the sample variance estimate for $\{y_i\}$, divided by $N$.

Now we can construct a suitable goodness-of-fit statistic for the CWT. Let us consider a
null hypothesis $Y(a,b)=Y_0(a,b)$ that we need to test. After that, construct the following
standardized quantity:
\begin{equation}
z(a,b) = \frac{\widetilde Y(a,b) - Y_0(a,b)}{\sqrt{\widetilde D(a,b)}}
\label{zdef}
\end{equation}
In fact, this $z(a,b)$ is nothing more than just the Student t-statistic for the sample
$\{y_i\}$, with an exception that the Student test was originally designed for strictly
Gaussian input data, while our $y_i$ are non-Gaussian in general. The statistic $z(a,b)$ is
asymptotically standard normal for $N\to\infty$, if $Y_0$ is true.

We adopt the test statistic~(\ref{zdef}) as the core of our method. Large values of
$|z(a,b)|$ point out some `structural patterns' in the p.d.f. $f(x)$, in addition to those
already included in the null model $Y_0$. It is legal to start the procedure from the
trivial null hypothesis $Y_0(a,b)\equiv 0$, if no initial information about $f(x)$ and
$Y(a,b)$ is available.

\section{Characterizing the noise level and calibrating the p-values}
\label{sec_noise}
\subsection{The problem}
The estimates~(\ref{wavest}) and~(\ref{zdef}) obviously contain random noise owing to the
discrete and finite nature of the sample $\{x_i\}$. This is different from the traditional
additive noise that appears in the time series analysis, so we cannot use any theoretic
results from this domain. The noise in our task is basically a shot noise instead. To
complete our statistical testing scheme, we must characterize the noise levels in $z(a,b)$
by determining its statistical distributions.

We adopt the frequentiest hypothesis testing framework and aim to compute the associated
p-value or `false alarm probability' (FAP hereafter). We want to test the following
hypotheses:
\begin{align}
H_0 &: Y(a,b) = Y_0(a,b) \text{ everywhere in } \mathcal D, \nonumber\\
H_A &: Y(a,b) \neq Y_0(a,b) \text{ somewhere in } \mathcal D,
\end{align}
where $\mathcal D$ is some predefined domain in the $(a,b)$ plane. The model $Y_0$ is
treated acceptable whenever $|z(a,b)|$ remains low within the domain $\mathcal D$:
\begin{equation}
|z(a,b)| < z_{\rm thr} \quad \forall (a,b) \in \mathcal D,
\end{equation}
which can be equivalently rewritten as
\begin{equation}
z_{\max} < z_{\rm thr}, \quad z_{\max} = \max_{(a,b)\in \mathcal D} |z(a,b)|.
\label{zthrmax}
\end{equation}
Now, $z_{\max}$ can be adopted as our test statistic: reject $H_0$ if $z_{\rm max}>z_{\rm
thr}$, and keep $H_0$ otherwise. Assuming that $H_0$ is true, define the distribution of
$z_{\rm max}$:
\begin{equation}
P_{\max}(z) = \Pr( z_{\max}<z | H_0 ).
\label{evd}
\end{equation}
We can derive the threshold level $z_{\rm thr}$ from the desired FAP as
$\FAP=1-P_{\max}(z_{\rm thr})$. However, to do this we must know the function $P_{\rm
max}(z)$, which is an extreme-value distribution (EVD) of the random field $z(a,b)$.
Analytic characterization of $P_{\rm max}(z)$ is the main problem treated in this section
below.

Our significance test is similar in several aspects to the method by \citet{Skuljan99}. The
crucial difference from \citet{Skuljan99} is that they consider different values
$\widetilde Y(a,b)$ \emph{individually}, while we treat them as an ensemble and construct
the \emph{maximized} statistic $z_{\max}$. As such, \citet{Skuljan99} deal with the
single-value distributions (SVD) of $\tilde Y(a,b)$, instead of the EVD~(\ref{evd}).
However, the SVD is not appropriate in this task, unless we know \emph{a priori} the exact
location $(a_0,b_0)$, where $\widetilde Y$ may possibly deviate from $Y_0$. In practice we
consider $\widetilde Y$ in a more or less wide domain $\mathcal D$ with no or little prior
knowledge on $(a,b)$, so we perform multiple single-value tests at once, increasing the
probability to admit a false alarm. The same effect appears in the periodogram time-series
analysis due to the unknown signal frequency, where it is called the `bandwidth penalty'
\citep[e.g.][]{HorneBal86,SchwCzerny98b,Baluev08a}. In our present task this effect may be
called the `domain penalty'.

To put our approach in a more general mathematical framework, the maximum modulo is an
$L_\infty$ metric. That is, we utilize the $L_\infty$ norm of $z(a,b)$ as an aggregate test
statistic. Though we do not consider other options below, this is definitely not the unique
treatment available in this task. In particular, we could use the $L_2$ norm of $z(a,b)$,
and it would lead us to a chi-square-like test. An approach of this kind was used by e.g.
\citet{McEwen04} to analyse the WMAP CMB data. We nonetheless prefer the approach based on
the extreme-value testing, since it allows us to easily identify which particular wavelet
coefficients (values of the $z$ statistic) appeared significant. In the case of the
$\chi^2$ statistic, it becomes unclear which points in the $(a,b)$ plane are responsible
for the observed $\chi^2$ deviation: just a narrow domain corresponding to the maximum peak
of $z(a,b)$, or maybe this includes some additional points, even though below the maximum,
contribute with a large amount due to a large cumulative area they occupy.

Another potentially promising approach might be offered by the $L_1$ metric minimization
and the compressed sensing techniques \citep{Hara17}.

\subsection{Noise distribution in the normal case}
The random field $z(a,b)$ is asymptotically Gaussian with standardized characteristics:
$\expect z(a,b)\to 0$ and $\disp z(a,b) \to 1$ for $N\to\infty$. Therefore, we should first
derive the FAP estimate for the limiting case, assuming that $N$ is large enough, so that
$z(a,b)$ is nearly standard Gaussian in the domain $\mathcal D$. EVDs for a standard
Gaussian random field were investigated in multiple works \citep[e.g.][for a
review]{AzaisWschebor-levelsets}. Our task falls under conditions of the generalized Rice
method \citep{AzaisDelmas02}, and we rely on this theory to derive the necessary FAP
estimation. An overview of the problem from a more practical point of view, explaning its
role in periodogram data analysis, was given in \citep{Baluev13b}. Our current task lead us
to the same abstract EVD problem.

We are going to apply the formulae of \citep[Sect.~3]{Baluev13b} to construct a FAP
estimate $\FAP=1-P_{\max}(z)$, which is a tail probability of $z_{\rm max}$. We try to keep
most of the notations here the same or at least similar. But the reader is cautioned about
a potentially misleading difference that $z$ in \citep{Baluev13b} would have the meaning of
$z^2/2$ here, while $Z$ from \citep{Baluev13b} is the same as our present $z$. We will not
discuss or explain the theory of the Rice method below, referring the reader to the cited
literature.

We should start from the one-sided version of~(\ref{zthrmax},\ref{evd}):
\begin{align}
\FAP^+(z_{\rm thr}) &= \Pr(z_{\max}^+>z_{\rm thr}), \nonumber\\
z_{\max}^+ &= \max_{(a,b)\in \mathcal D} z(a,b),
\label{evdp}
\end{align}
where we maximize $z$ itself rather than its absolute value $|z|$. We use only the primary
term in the $\FAP$ approximation, given by the generalized Rice formula:
\begin{equation}
\FAP^+(z_{\rm thr}) \sim \int\limits_{\mathcal D} da db \int\limits_{z_{\rm thr}}^{+\infty} dz \int\limits_{\mathbb R^3} p_{zz'z''}(z,0,z'') \det z'' dz''.
\label{Rice}
\end{equation}
The integral in~(\ref{Rice}) gives the average number of local maxima of $z(a,b)$ inside
$\mathcal D$ that rise above the threshold $z_{\rm thr}$. This is an
asymptotic approximation valid for $z_{\rm thr}\to\infty$, or $\FAP\to 0$.
In particular, we neglected the effect of the boundary of $\mathcal D$
that has the relative magnitude of $\sim 1/z_{\rm thr}$. In~(\ref{Rice}), the notation
$p_{zz'z''}$ stands for the $6$-variate p.d.f. of the random field value $z$, its gradient
$z'$, and its Hessian matrix $z''$, if all are computed at the given point
$(a,b)$.

For a standard Gaussian $z(a,b)$, the integral in~(\ref{Rice}) becomes:
\begin{equation}
\FAP^+(z) \sim A_0 \left(\sqrt{\frac{2}{\pi}} z e^{-z^2/2} + \erfc \frac{z}{\sqrt 2} \right) + A_1 \erfc \frac{z}{\sqrt 2},
\label{fap1}
\end{equation}
This is a special case of formula~(20) from \citep{Baluev13b}. We rewrite it as:
\begin{align}
\FAP^+(z) &\sim W_{00} z e^{-z^2/2} + W_{02} \sqrt{\frac{\pi}{2}} \erfc \frac{z}{\sqrt 2} \nonumber\\
          &\sim \left( W_{00} + W_{02} z^{-2} + \ldots \right) z e^{-z^2/2},
\label{FAPgauss}
\end{align}
where $W_{00}$ and $W_{02}$ are coefficients that could be expressed via $A_{0,1}$. We
follow the convention that the index $j$ in $W_{ij}$ corresponds to the power $z^{-j}$
in the relevant term of the expansion. That is, the term with $W_{02}$ has the
relative magnitude of $\sim 1/z^2$ and hence can be neglected for large $z$.
Additional index $i$ is reserved for non-Gaussian corrections to~(\ref{FAPgauss}) that will
be considered below.

The primary coefficeint $W_{00}$ becomes
\begin{equation}
W_{00} = \frac{1}{(2\pi)^\frac{3}{2}} \int\limits_{\mathcal D} \sqrt{\det \mathbmss G(a,b)}\, da db, \quad \mathbmss G = \var z',
\label{W00}
\end{equation}
where the $2\times 2$ matrix $\mathbmss G$ is the covariance matrix of the gradient $z'$.

So far in this subsection we performed mainly repacking and adaptation of the formulae from
\citep{Baluev13b}. But now the covariance matrix $\mathbmss G$ in~(\ref{W00}) depends on
specific characteristics of our random field. The calculation of $\mathbmss G$
involves relatively long but routine manipulations. We do this by applying the following
sequence:
\begin{enumerate}
\item compute the gradient of~(\ref{zdef});
\item average its pairwise products with respect to $x_i$ (using $f(x)$ as the p.d.f.);
\item whenever legitimate, replace various sample momenta by their
integral counterparts, e.g. $\langle y^2\rangle -\langle y \rangle^2
\to \disp y$ for large $N$.
\end{enumerate}

The matrix $\mathbmss G$ is finally expressed as:
\begin{equation}
\lim_{N\to\infty} G_{ij} = \frac{\cov(y'_i,y'_j)}{\disp y} - \frac{\cov(y,y'_i)\cov(y,y'_j)}{(\disp y)^2},
\label{zgvar}
\end{equation}
where an index near derivatives denotes the variable, either $a$ or $b$. As we can
see, $\mathbmss G$ depends on the wavelet $\psi$ (via $y$), its first derivative $\psi'$
(via $y'$), and also on the $x$-distribution p.d.f. $f(x)$ via the covariance momenta. The
latter dependence is particularly troubling, because we do not know $f(x)$ at this
stage: its characterization is actually our final goal. However, instead
of using~(\ref{zgvar}) directly, we may substitute an estimate in its place. In particular,
it is convenient to replace all covariances in~(\ref{zgvar}) by their sample estimates:
\begin{align}
\cov(y'_i,y'_j) &\simeq \langle y'_i y'_j \rangle - \langle y'_i \rangle \langle y'_j \rangle, \nonumber\\
\cov(y,y'_i) &\simeq \langle y y'_i \rangle - \langle y \rangle \langle y'_i \rangle, \nonumber\\
\disp y &\simeq \langle y^2 \rangle - \langle y \rangle^2.
\label{covest}
\end{align}
Such a replace yields an estimate $\widetilde{\mathbmss G}$ that we can substitute
in~(\ref{W00}) instead of $\mathbmss G$ and then integrate it numerically. This infers a
relative error of $\sim 1/\sqrt N$ in the result.

Finally, let us consider the two-sided estimation $\FAP$ (instead of the one-sided
$\FAP^+$). To treat this case we must double~(\ref{Rice}), because the statistics of
positive and negative maxima is symmetric and their average numbers must be identical.
Therefore, we may just double~(\ref{FAPgauss}):
\begin{equation}
\FAP(z) \sim 2 W_{00} z e^{-z^2/2} + W_{02} \sqrt{2\pi} \erfc \frac{z}{\sqrt 2}.
\label{dsFAPgauss}
\end{equation}

\subsection{Handling deviations from the normality}
Non-Gaussian deviations in the $z$-distribution may constitute a serious issue. It concerns
more deep matters than just the accuracy of the $\FAP$ estimate~(\ref{FAPgauss}).
If~(\ref{FAPgauss}) is by any reason inaccurate then we could just perform Monte Carlo
simulations to acquire a better estimate. But non-Gaussianity also results in statistical
asymmetry between positive and negative values of $z(a,b)$, and non-Gaussian deviations
vary over the $(a,b)$ plane. This may cause an inhomogeneous noise in the SWT, when only a
minor fraction of points supply a dominant contribution to the extreme-value statistic
$z_{\max}$. This is a pathological situation that should be avoided
\citep[sect.~5.8]{Jansen-NRWT}.

Consider the scale $a$ is small. Then most $y_i$ terms in the sum~(\ref{wavest}) become
negligible, because most of $x_i$ fall outside of the wavelet localization range. This has
an effect of decreasing the `effective $N$'. In the ultimate case, just a few of $x_i$
contribute significant amount in the sum~(\ref{wavest}), rendering its distribution
drastically non-Gaussian. Therefore, our analysis is meaningful only if it does not
consider too small $a$. Depending on $N$, this small-scale limit may be larger or smaller,
and we must find a way to determine it.

Roughly, the effective number of terms contributing in the sum~(\ref{wavest}) can be
estimated using the following formula:
\begin{equation}
n(a,b) = \left[ \sum_{i=1}^N \varphi\left(\frac{x_i-b}{a}\right) \right] \left/ \int\limits_{-\infty}^{+\infty} \varphi(x) dx \right. .
\label{num}
\end{equation}
Since $\varphi$ has approximately the same localization as $\psi$, the number of terms
dominating in~(\ref{num}) is approximately the same as in~(\ref{wavest}), but now all these
terms are positive and never cancel each other.

The characteristic~(\ref{num}) can be used as a simple criterion of the normality, but it
appears too rough. Simulations revealed that with some wavelets we need at least $n=100$ to
achieve a good normality, while others yield satisfactory results already for $n=10$.
Obviously, there is more deep dependence on the wavelet shape, and we should seek more
subtle normality criteria than~(\ref{num}).

\citet{Martins10} investigated the Student $t$-statistic like our~(\ref{zdef}) for
non-Gaussian data and expressed its first momenta. Using these results, we can write down
the following characteristics:
\begin{align}
\expect z &\simeq -\frac{1}{2} \frac{\As y}{\sqrt N}, \nonumber\\
\disp z &\simeq 1 + \frac{1}{N} \left[2+\frac{7}{4} (\As y)^2 \right], \nonumber\\
\As z &\simeq -2 \frac{\As y}{\sqrt N}, \nonumber\\
\Ex z &\simeq \frac{1}{N} \left[ 12 (\As y)^2 - 2 \Ex y + 6 \right],
\label{zmom}
\end{align}
where $\As$ is skewness and $\Ex$ is excess kurtosis. As we can see, the skewness $\As y$
plays a dominant role here. Interestingly, we also obtain some deviations in the average
and variance of $z$, i.e. not only the normality of $z$ is disturbed, but also its
normalization gets broken and a bias appears. It follows from~(\ref{zmom}) that $\As y$ and
$\Ex y$ could be used to construct the dedicated normality test. However, it is unclear at
this stage, how to formulate such a criterion, because there is no connection
between~(\ref{zmom}) and our $\FAP$ estimate~(\ref{FAPgauss}).

To establish this connection, we employ a series decomposition of the Edgeworth type. For
the p.d.f. of the $z$-statistics it looks like:
\begin{align}
p_{\rm nongauss}(z) &\simeq p_{\rm gauss}(z) + \frac{1}{\sqrt N} C_{11} p_{11}(z) + \nonumber\\
&\quad + \frac{1}{N} [C_{21} p_{21}(z) + C_{22} p_{22}(z)] + \ldots,
\label{edge}
\end{align}
where the coefficients $C_{ij}$ can be expressed via the momenta~(\ref{zmom}), and $p_i(z)$
are derivatives of $p_{\rm gauss}(z)$. We constructed a similar multivariate series for the
$6$-dim p.d.f. $p_{zz'z''}$ that appears in the Rice formula~(\ref{Rice}), and then
integrate it term-by-term. The result is an Edgeworth-type series for the $\FAP$.

We cannot provide this computation in full details, because some decompositions appearing
on the way contained up to $\sim 60000$ terms. We used a computer algebra system to reach
the goal, and a skeleton of this computation is provided in \ref{sec_edge}, while the MAPLE
worksheet containing these computations is attached as the online-only material (see
\ref{sec_maple}). This computation resulted in the following double-series expansion that
extends~(\ref{FAPgauss}) to the non-Gaussian case:
\begin{align}
\FAP^\pm(z) &\sim z e^{-z^2} \times \left[ \vphantom{\frac{1}{N}} \left(W_{00} + W_{02} z^{-2} + \ldots \right) \pm \right. \nonumber\\
&\pm \frac{1}{\sqrt N}\left( W_{1,-3} z^3 + W_{1,-1} z + \ldots \right) + \nonumber\\
&\left. + \frac{1}{N}\left( W_{2,-6} z^6 + W_{2,-4} z^4 + \ldots \right) + \ldots \right].
\label{FAPngauss}
\end{align}

We did not plan to use the decomposition~(\ref{FAPngauss}) in the direct way, i.e. to
compute non-Gaussian corrections to the p-values. In fact, it looks likely that whenever
the non-Gaussian deviations are significant indeed, this decomposition may have poor
convergence, if it converges at all. But we only need to determine the domain in the
$(a,b)$ plane, where $z(a,b)$ guaranteedly preserve almost Gaussian behaviour, enabling us
using~(\ref{FAPgauss}) without any corrections at all. This opportunity reveals itself if
we look at the general expression for $W_{ij}$:
\begin{equation}
W_{ij} = \frac{1}{(2\pi)^\frac{3}{2}} \int\limits_{\mathcal D} q_{ij}(a,b) \sqrt{\det \mathbmss G(a,b)}\, da db.
\label{Wij}
\end{equation}
It is very similar to~(\ref{W00}) though contains an additional multiplier $q_{ij}$ in the
integrand. Obviously, if $q_{ij}(a,b)$ is small, the
particular point $(a,b)$ contributes with a relatively small amount to the
selected coefficient $W_{ij}$, compared to its contribution in the primary coefficient
$W_{00}$. On contrary, if we consider only points with small $q_{ij}$, the
integral~(\ref{Wij}) should appear small relatively to $W_{00}$, and hence the non-Gaussian
deviations should remain small too.

Therefore, our normality criterion may involve a combination of $4$ quantities: $q_{1,-3}$,
$q_{1,-1}$, $q_{2,-6}$, and $q_{2,-4}$ (while $q_{02}$ is unrelated to non-Gaussianity).
These $q_{ij}$ should be multiplied by some powers of $z$ and $N$ in accordance with the
relevant terms of~(\ref{FAPngauss}). After that, we join them together in a sum-of-squares
metric and construct the following criterion:
\begin{equation}
\frac{1}{N} \left( q_{1,-3}^2 z_*^6 + q_{1,-1}^2 z_*^2 \right) +
\frac{1}{N^2} \left( q_{2,-6}^2 z_*^{12} + q_{2,-4}^2 z_*^8 \right) < \varepsilon^2
\label{nrmtest}
\end{equation}
Note that the normality depends on the $z$ level (larger $z$, less Gaussian), so we had to
specify a characteristic $z_*$ in~(\ref{nrmtest}). The meaning of this $z_*$ is the maximum
$z$-level, for which the normality is guaranteedly preserved. It depends on the desired
$\FAP$-level that we expect to approximate. In practice these characteristic $z$-levels do
not change much: they usually reside in the range $3-5$ for all reasonable values of
parameters. We employ a universal value $z_*^2=10$. Another control parameter
in~(\ref{nrmtest}) is $\varepsilon$, and it sets the desired limit on the relative
magnitude of possible non-Gaussian corrections. We usually set $\varepsilon^2=0.1$.

The computation sequence and all formulae necessary to compute the coefficients $q_{ij}$
are given in \ref{sec_edge}. Reviewing them, we can confirm that the skewness $\As y$ is
solely responsible for the greatest terms in~(\ref{nrmtest}), $q_{1,-3}$ and $q_{2,-6}$.
The normality domain would enlarge significantly, if $\As y$ is somehow suppressed.

When applying~(\ref{nrmtest}) in practice, we noticed that the normality can be broken not
only in the small-scale range, as we expected, but also on the opposite, large-scale side
of the $(a,b)$ plane (if $a$ comparable to the sample variance and above). However, if the
small-scale deviation indicated, basically, a developing degeneracy of the
$z$-distribution, the non-Gaussianity in the large-scale range is not such a serious issue.
This is an effect of a partial coverage of a highly dilated wavelet, and it does not
necessarily infer a degeneracy in the $z$-distribution. When $a$ is large, the values of
$z(a,b)$ are typically very large too, and hence undoubtfully significant, even though
non-Gaussian. Simultaneously the large-scale part of the CWT is important in recovering the
correct normalization of $f(x)$ (as a p.d.f., it should always integrate to unit).
Therefore, in the large-scale domain $a\gtrsim \sigma_x$ it is reasonable to replace the
full criterion~(\ref{nrmtest}) by the simplified one~(\ref{num}). In such a way,
non-Gaussian points at large scales are removed from the analysis, only if the associated
number statistics appeared too poor.

\section{Finding optimal wavelets}
\label{sec_opt}
\subsection{Motivation}
So far we paid relatively little attention to the choice of the wavelet $\psi(t)$. We only
required it to be the first or second derivative of some bell-shaped generating function
$\varphi(t)$, suitable for smoothing. The requirement of having the `bell-like shape' can
be formalized as follows:
\begin{eqnarray}
\varphi(t)>0 \; {\rm everywhere}, \quad \varphi(t)=\varphi(-t), \nonumber\\
\varphi'(t)>0 \; {\rm for} \; t<0, \quad \varphi'(t)<0 \; {\rm for} \; t>0.
\label{gfc}
\end{eqnarray}
That is, $\varphi(t)$ must be even and always positive, and must have a single local
maximum at $t=0$, monotonically decreasing in the tails.

We still have much freedom in selecting $\varphi$. We may seek such a $\varphi$ that would
improve some properties of our analysis algorithm: reduce the noise level or
reduce normality deviations in $z$.

\subsection{Wavelets normalization}
Before proceeding further, we need to address a normalization issue for $\varphi(t)$.
Having some initial $\varphi(t)$, we can arbitrarily rescale it to $\varphi_1(t) =
K\varphi(kt)$. This new generating function is `physically' identical to the initial one,
nut nonetheless $\varphi_1$ and $\varphi$ differ in many their formal characteristics. It
is convienient to introduce some standard normalization of $\varphi(t)$, and compare two
generating functions or two wavelets only after they both are properly rescaled.

The first natural normalizing condition is that $\varphi$ must be a weight function, that
is it must integrate to $1$. When such a $\varphi$ is used for smoothing, it maps a
constant function to the same constant. This enables us to treat our analysis intuitively,
directly comparing the values of $Y$ with $f'$ or $f''$.

The second condition is intended to normalize the noise in the wavelet transform,
expression~(\ref{wavvar}). Of course, achieving a strictly constant $D(a,b)$ is impossible,
but we can reach this in the most important small-scale region. For small $a$, $f(x)$ can
be treated almost constant inside the wavelet localization segment. Therefore, we can
simplify~(\ref{wavdef}) and~(\ref{wavvar}) to
\begin{align}
Y &= a \int\limits_{-\infty}^{+\infty} f(b+at) \psi(t) dt \approx a f(b) \int\limits_{-\infty}^{+\infty} \psi(t) dt =0, \nonumber\\
D &= \frac{1}{N} \left[ a \int\limits_{-\infty}^{+\infty} f(b+at) \psi^2(t) dt - Y^2 \right] \approx \nonumber\\
 &\quad \approx \frac{a f(b)}{N} \int\limits_{-\infty}^{+\infty} \psi^2(t) dt.
\label{YDss}
\end{align}
Hence, to equibalance $D(a,b)$ for different wavelets $\psi$, all these wavelets must have
the same $L_2$ norm.

Summarizing, our two normalization condition can be written down as follows:
\begin{equation}
\int\limits_{-\infty}^{+\infty} \varphi(t) dt =1, \qquad \int\limits_{-\infty}^{+\infty} \psi^2(t) dt =1.
\label{wavnrm}
\end{equation}

Now, assume that we have some unnormalized $\varphi(t)$ and seek such coefficients $k$ and
$K$ that $\varphi_{\rm nrm}(t)=K \varphi(k t)$ satisfies~(\ref{wavnrm}). In the first case,
$\psi=-\varphi'$, we obtain:
\begin{align}
\varphi_{\rm nrm}(t) &= K\varphi(kt), \nonumber\\
\psi_{\rm nrm}(t) &= -Kk\varphi'(kt), \nonumber\\
K &= \left(\int\limits_{-\infty}^{+\infty} \varphi\, dt \int\limits_{-\infty}^{+\infty} \varphi'^2 dt \right)^{-\frac{1}{3}}, \nonumber\\
k &= \left(\int\limits_{-\infty}^{+\infty} \varphi\, dt \right)^{\frac{2}{3}} \left(\int\limits_{-\infty}^{+\infty} \varphi'^2 dt \right)^{-\frac{1}{3}}.
\label{oddnrm}
\end{align}
In the second case, $\psi=\varphi''$, we have
\begin{align}
\varphi_{\rm nrm}(t) &= K\varphi(kt), \nonumber\\
\psi_{\rm nrm}(t) &= Kk^2\varphi''(kt), \nonumber\\
K &= \left(\int\limits_{-\infty}^{+\infty} \varphi\, dt\right)^{-\frac{3}{5}} \left(\int\limits_{-\infty}^{+\infty} \varphi''^2 dt \right)^{-\frac{1}{5}}, \nonumber\\
k &= \left(\int\limits_{-\infty}^{+\infty} \varphi\, dt \right)^{\frac{2}{5}} \left(\int\limits_{-\infty}^{+\infty} \varphi''^2 dt \right)^{-\frac{1}{5}}.
\label{evennrm}
\end{align}
For examaple, for~(\ref{wHerm}):
\begin{align}
\text{WAVE} &: K=0.7663992, \; k=1.921078, \nonumber\\
\text{MHAT} &: K=0.5442755, \; k=1.364296.
\end{align}

\subsection{Formalizing wavelet optimality}
We want to improve our wavelet in two aspects: (i) improve the normality of $z(a,b)$, and
(ii) improve the sensitivity of $z(a,b)$ to weak signals. Let us consider these tasks
individually.

As we noticed in Sect.~\ref{sec_noise}, the skewness $\As y$ is responsible for the
largest-order terms in the normality criterion~(\ref{nrmtest}). Therefore, we must find
such a wavelet that has $\As y=0$ if possible. In general, this goal cannot be achieved in
the entire $(a,b)$ plane, but we may try to do this for small $a$, where the
non-Gaussianity is more important. In this case we obtain
\begin{equation}
\As y = \frac{\expect (y-Y)^3}{(\disp y)^{3/2}} \approx \frac{1}{\sqrt{a f(b)}} \frac{\int\limits_{-\infty}^{+\infty} \psi^3(t) dt}{\left(\int\limits_{-\infty}^{+\infty} \psi^2(t) dt \right)^{3/2}}.
\end{equation}
Hence, $\As y$ would (almost) vanish if the wavelet satisfies the restriction
\begin{equation}
\int\limits_{-\infty}^{+\infty} \psi^3(t) dt = 0.
\label{psi3}
\end{equation}
If $\psi=-\varphi'$ then $\psi$ is an odd function. Equation~(\ref{psi3}) is always
satisfied for odd wavelets. Deviations from the normality should be small for such wavelets
`as is', without making any special efforts. On contrary, if $\psi=\varphi''$ then it is an
even function, and the equality~(\ref{psi3}) becomes non-trivial. In particular, the MHAT
wavelet does not satisfy~(\ref{psi3}) and generates large skewness effects in $z(a,b)$.

Now let us consider improving the sensitivity of $z(a,b)$. In the FAP
approximation~(\ref{FAPgauss}), larger $W_{00}$ imply smaller significance and larger
noise. Therefore, we should try to minimize $W_{00}$ by seeking an optimal wavelet $\psi$.
We must also take care of the normalization~(\ref{wavnrm}), because otherwise we may reduce
noise level simultaneously with the signal (for example, if we merely scale $\psi$ down),
or just move the noise to smaller $a$ without changing its magnitude.

To minimize $W_{00}$, we apply the small-scale approximation to~(\ref{zgvar}), taking into
account the normalization~(\ref{wavnrm}). After easy manipulations, this yields
\begin{equation}
a^4 \det \mathbmss G \approx T(\psi) = \left(\int\limits_{-\infty}^{+\infty} t^2 \psi'^2 dt - \frac{1}{4}\right) \int\limits_{-\infty}^{+\infty} \psi'^2 dt.
\label{detGss}
\end{equation}
According to~(\ref{W00}) and~(\ref{detGss}), the magintude of $W_{00}$ is proportional to
$T(\psi)$. Therefore, minimizing $W_{00}$ is equivalent to minimizing the functional
$T(\psi)$ with respect to the wavelet $\psi$, taking into account the normalization
constraints~(\ref{wavnrm}).

\subsection{Finding optimal solutions}
To solve the variational task of the previous subsection, we pre-normalize $\psi$
using~(\ref{oddnrm},\ref{evennrm}) and then substitute $\psi_{\rm nrm}$ to~(\ref{detGss}).
For odd wavelets we obtained
\begin{align}
T_{\rm nrm}(\varphi) &= \left(\int\limits_{-\infty}^{+\infty} t^2 \varphi''^2 dt - \frac{1}{4} \int\limits_{-\infty}^{+\infty} \varphi'^2 dt\right)
\left(\int\limits_{-\infty}^{+\infty} \varphi''^2 dt \right) \times \nonumber\\
&\quad \times \left(\int\limits_{-\infty}^{+\infty} \varphi'^2 dt \right)^{-\frac{8}{3}}
\left(\int\limits_{-\infty}^{+\infty} \varphi\, dt \right)^{\frac{4}{3}},
\label{Tnrmodd}
\end{align}
while in the even case
\begin{align}
T_{\rm nrm}(\varphi) &= \left(\int\limits_{-\infty}^{+\infty} t^2 \varphi'''^2 dt - \frac{1}{4}\int\limits_{-\infty}^{+\infty} \varphi''^2 dt\right)
\left(\int\limits_{-\infty}^{+\infty} \varphi'''^2 dt \right) \times \nonumber\\
&\quad \times \left(\int\limits_{-\infty}^{+\infty} \varphi''^2 dt\right)^{-\frac{12}{5}}
\left(\int\limits_{-\infty}^{+\infty} \varphi\, dt \right)^{\frac{4}{5}}.
\label{Tnrmeven}
\end{align}
Contrary to $T$, the appropriate $T_{\rm nrm}$ can be minimized without paying attention to
the constraints~(\ref{wavnrm}). In the even case, the skewness elimination
constraint~(\ref{psi3}) must be taken into account in this optimization. And in any case,
the general requirements~(\ref{gfc}) must be satisfied. Note that there is a trivial
minimum $T_{\rm nrm}=0$, achieved if $\varphi$ integrates to zero, but such a generating
function violates~(\ref{gfc}) and cannot be used as a smoothing kernel.

The task appears too difficult with arbitrary $\varphi$, so we first adopt a simple
parametric model for it and then perform the optimization with respect to the model
parameters. We set the following model:
\begin{align}
\varphi(t) &= P(t^2) e^{-\frac{t^2}{2}}, &P(u) &= p_0 + p_1 u + \ldots + p_m u^m, \nonumber\\
\psi_1(t) &= P_1(t^2) t e^{-\frac{t^2}{2}}, &P_1(u) &= P(u) - 2 P'(u), \nonumber\\
\psi_2(t) &= P_2(t^2) e^{-\frac{t^2}{2}}, &P_2(u) &= (u-1) P_1(u) - 2u P_1'(u).
\label{wmp}
\end{align}
The optimization is made by varying the coefficients $p_i$. Since the scale factor is
arbitrary here, we set $p_m=1$ and treat other $p_i$ as free.

Substituting~(\ref{wmp}) to the objective~(\ref{detGss}), as well as to the
constraints~(\ref{wavnrm}) and~(\ref{psi3}), yields some multivariate polynomials depending
on $p_i$. Therefore, our optimization task is reduced to a system of algebraic equations.
Such systems can be solved by reducing them to a single polynomial equation of a higher
degree, with respect to a single unknown variable, some of $p_i$ in our case. Each root of
the latter polynomial generates one possible solution to the system, and by finding all
real roots we may obtain the whole set of solutions. After that, we need to select only
those solutions that satisfy~(\ref{gfc}), and also conider the relevant boundary minima.
These constraints can also be formulized via algebraic restrictions.

\begin{figure*}[!t]
\includegraphics[width=\textwidth]{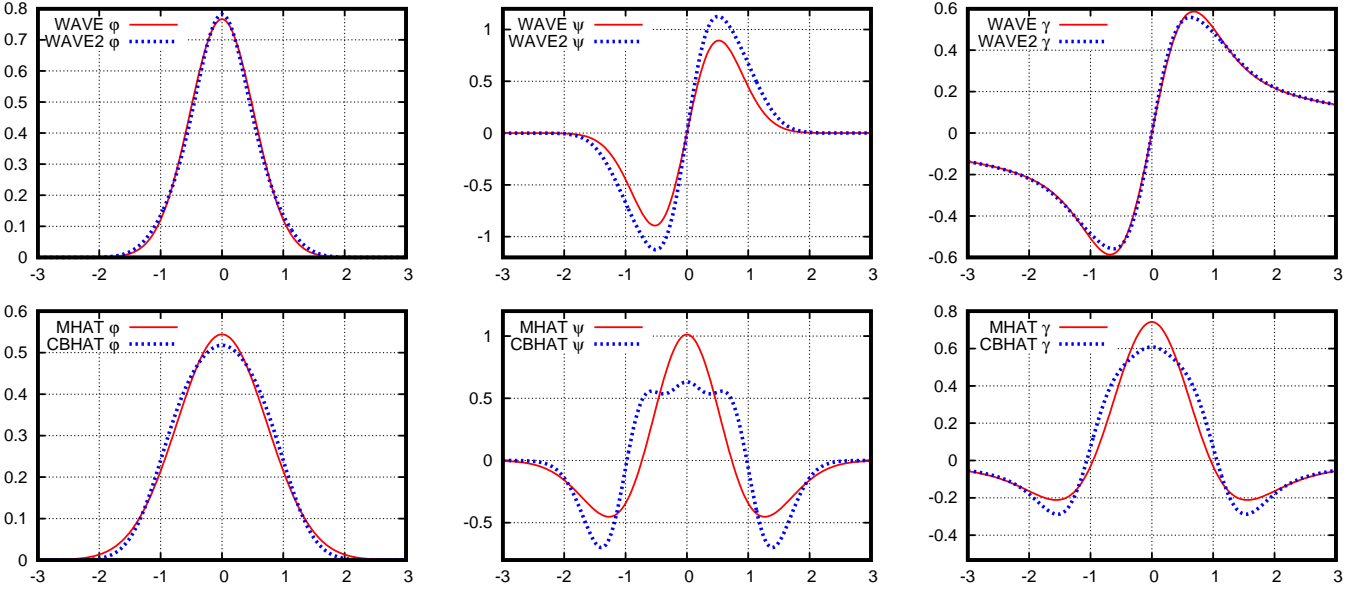}
\caption{Different analysing wavelets $\psi$, their generating functions $\varphi$, and the
associated optimal reconstruction kernels $\gamma$. Everything after the
normalization~(\ref{wavnrm}).}
\label{fig_wavs}
\end{figure*}

The degree of the polynomials arising on the way is pretty large, so we seeked the help
from the computer algebra again, and now give only a summary of the results. We considered
quadratic polynomials $P(u)$, $m=2$, and found the following:
\begin{enumerate}
\item For odd wavelets there is a single solution satisfying~(\ref{gfc}) that we called
WAVE2, with $T_{\rm nrm}$ slightly smaller than for the WAVE wavelet ($T_{\rm nrm}=0.170$
against $T_{\rm nrm}=0.183$). This WAVE2 wavelet is very close to WAVE
(Fig.~\ref{fig_wavs}).
\item For even wavelets we found a single local minimum of $T_{\rm nrm}$
satisfying~(\ref{gfc}). This minimum corresponds to a suitable optimal solution with
$T_{\rm nrm}=0.479$. We called this new wavelet CBHAT, or `Cowboy Hat', by its reminiscent
shape~(Fig.~\ref{fig_wavs}). Note that the classic
MHAT wavelet has a smaller $T_{\rm nrm}=0.217$, but this is by the cost of
breaking~(\ref{psi3}) and largerly skewed noise in CWT. We put more priority on suppressing
the non-Gaussianity, and therefore we adopt CBHAT wavelet as a better replacement of MHAT.
\item Boundary optima did not allow to further reduce $T_{\rm nrm}$. All optimal solutions
of this task are located strictly inside the domain~(\ref{gfc}).
\end{enumerate}

Our new optimal wavelets are defined by the formulae
\begin{align}
\varphi(t) &= (24.8929 + 0.3794 t^2 + t^4) e^{-\frac{t^2}{2}}, \nonumber\\
\psi_1(t) &= (24.1342 - 3.6206 t^2 + t^4) t e^{-\frac{t^2}{2}}, \nonumber\\
&K=0.0313959, \; k=2.22497
\label{WAVE2}
\end{align}
for WAVE2, and like
\begin{align}
\varphi(t) &= (14.9952 + 5.2378 t^2 + t^4) e^{-\frac{t^2}{2}}, \nonumber\\
\psi_2(t) &= (-4.5196 + 0.8062 t^2 - 3.7622 t^4 + t^6) e^{-\frac{t^2}{2}}, \nonumber\\
&K=0.0345363, \; k=2.01127
\label{CBHAT}
\end{align}
for CBHAT. Their graphs are shown in Fig.~\ref{fig_wavs}. There is a noticable similarity
between the generating functions in the WAVE/WAVE2 and in the MHAT/CBHAT pairs. This means
that e.g. the MHAT and CBHAT CWTs should remain very similar to each other if the noise is
negligible. But the noise properties with CBHAT are much better than with MHAT. The odd
wavelets WAVE and WAVE2 are very close to each other, though WAVE2 offers slightly better
sensitivity.

\section{Wavelet reconstruction of the probability density function}
\label{sec_rec}
Basically, our $z$ statistic~(\ref{zdef}) together with the $\FAP$
estimate~(\ref{dsFAPgauss}) provide a goodness-of-fit test for the wavelet transform. We
know that if the null hypothesis $Y_0(a,b)$ was correct, the random field $z(a,b)$ should
likely stay within a $\pm z_{\rm thr}$ band about zero. Then a natural question appears:
what is the simplest p.d.f. model $f_0(x)$ that we must adopt to keep $z(a,b)$ solely
within this admissible noise band?

We formulate a rather simplified iterative method that tries to construct $f(x)$ from the
least possible number of the detected patterns. A single iteration involves: (i)
thresholding all values of $\widetilde Y$ that are consistent with $Y_0$, and (ii) applying
an inversion formula~(\ref{wavinv}) to this cleared wavelet transform. Such iterations can
be viewed to belong to the matching pursuit family, and the ``simplicity'' of $f(x)$ is
understood in terms of sparsity of its CWT.

Two methods, named as `hard' and `soft' thresholding, can be used \citep{ABS00}. In the
first case, the thresholded value $Y_{\rm thr}$ is set to either $Y_0$ or $\widetilde Y$,
depending on whether the statistic $z$ appeared significant or not. In the second case, the
value of $\widetilde Y$ is changed by the maximum admissible amount, $\pm z_{\rm thr}
\sqrt{\widetilde D}$, in order to shrink the absolute deviation $|Y_{\rm thr}-Y_0|$. In the
latter case, $Y_{\rm thr}(a,b)$ becomes continuous at the signal/noise transition. The
choice of the hard or soft thresholding scheme is a matter of the variance/bias tradeoff:
the `hard' version results in a larger variance, while the `soft' one has larger bias.

After thresholding, we should apply an inversion formula to $Y_{\rm thr}$ in order to
reconstruct $f(x)$. But to do this, we must define the reconstruction kernel $\gamma$. We
tried to find an optimal reconstruction kernel $\gamma$ based on the requirement to
minimize the random noise in the reconstructed $f(x)$. This task is considered in
\ref{sec_recon}, where the necessary optimal kernel $\gamma$ was determined via its Fourier
image as
\begin{equation}
\hat\gamma(\omega) \propto \hat\psi(\omega) / |\omega|.
\end{equation}
For example, for the WAVE wavelet we should set
\begin{equation}
\gamma_{\rm WAVE}(t) = \int\limits_0^\infty e^{-\frac{\omega^2}{2}} \sin\omega t\, d\omega = \sqrt 2\, \Dw\left(\frac{t}{\sqrt 2}\right),
\end{equation}
where $\Dw(x)$ is the Dawson function. This function looks generally similar to the WAVE
wavelet, with much heavier tails though ($\sim 1/x$ for large $x$). The reconstruction
kernel for the even case, $\psi_2=-\varphi''$, can be obtained by differentiating:
$\gamma_2(t)=-\gamma_1'(t)$, looks similar to MHAT (Fig.~\ref{fig_wavs}). Longer tails
result in a more smoothed reconstruction than for $\gamma=\psi$. Also, the side minima of
$\gamma_2$ are smaller than those of $\psi$, resulting in reduced side artifacts.

For the wavelet model~(\ref{wmp}) we derived the following general expressions:
\begin{align}
\gamma_1(t) &= p_0 d + p_1 t (td-1) + p_2 t (t^3d-1-t^2) + \ldots, \nonumber\\
\gamma_2(t) &= p_0 (td - 1) + p_1 [(t^2-2)td-1-t^2] + \nonumber\\
&\quad + p_2 [(t^2-4) t^3d+1+3t^2-t^4] + \ldots, \nonumber\\
&\quad d = \sqrt 2\, \Dw\left(\frac{t}{\sqrt 2}\right).
\label{gamma}
\end{align}
These formulae can be used to compute $\gamma_1$ for WAVE2 and $\gamma_2$ for CBHAT
wavelets by substituting the coefficients from~(\ref{WAVE2},\ref{CBHAT}). Their
normalization constants become
\begin{align}
C_{\psi_1\gamma_1} &= \pi^2\sqrt 2 \left(p_0^2 + p_0 p_1 + \frac{3}{2} p_0 p_2 + \frac{3}{4} p_1^2 + \right. \nonumber\\
&\quad \left. + \frac{15}{4} p_1 p_2 + \frac{105}{16} p_2^2 + \ldots \right), \nonumber\\
C_{\psi_2\gamma_2} &= \frac{\pi^2}{\sqrt 2} \left(p_0^2 - p_0 p_1 - \frac{9}{2} p_0 p_2 + \frac{7}{4} p_1^2 + \right. \nonumber\\
&\quad \left. + \frac{21}{4} p_1 p_2 + \frac{225}{16} p_2^2 + \ldots \right).
\end{align}

Whenever the wavelets are normalized according to~(\ref{oddnrm},\ref{evennrm}), their
reconstruction kernels scale as $K \gamma_1(kt)$ or $Kk\gamma_2(kt)$, while the constants
$C_{\psi\gamma}$ scale by $K^2/k$ and by $K^2k$, respectively. In particular,
\begin{align}
C_{\rm WAVE} &\approx 4.2676, &C_{\rm MHAT} &\approx 2.8205, \nonumber\\
C_{\rm WAVE2} &\approx 4.1710, &C_{\rm CBHAT} &\approx 2.8195.
\label{Cpg}
\end{align}

\section{Test simulations and the exoplanetary example}
\label{sec_simul}
Approximating the false alarm probability was one of the most complicated part of the
method, possibly vulnerable to mistakes. Therefore, we need to verify the relevant formulae
by comparing them with numerical simulations.

\begin{figure*}[!t]
\includegraphics[width=0.49\textwidth]{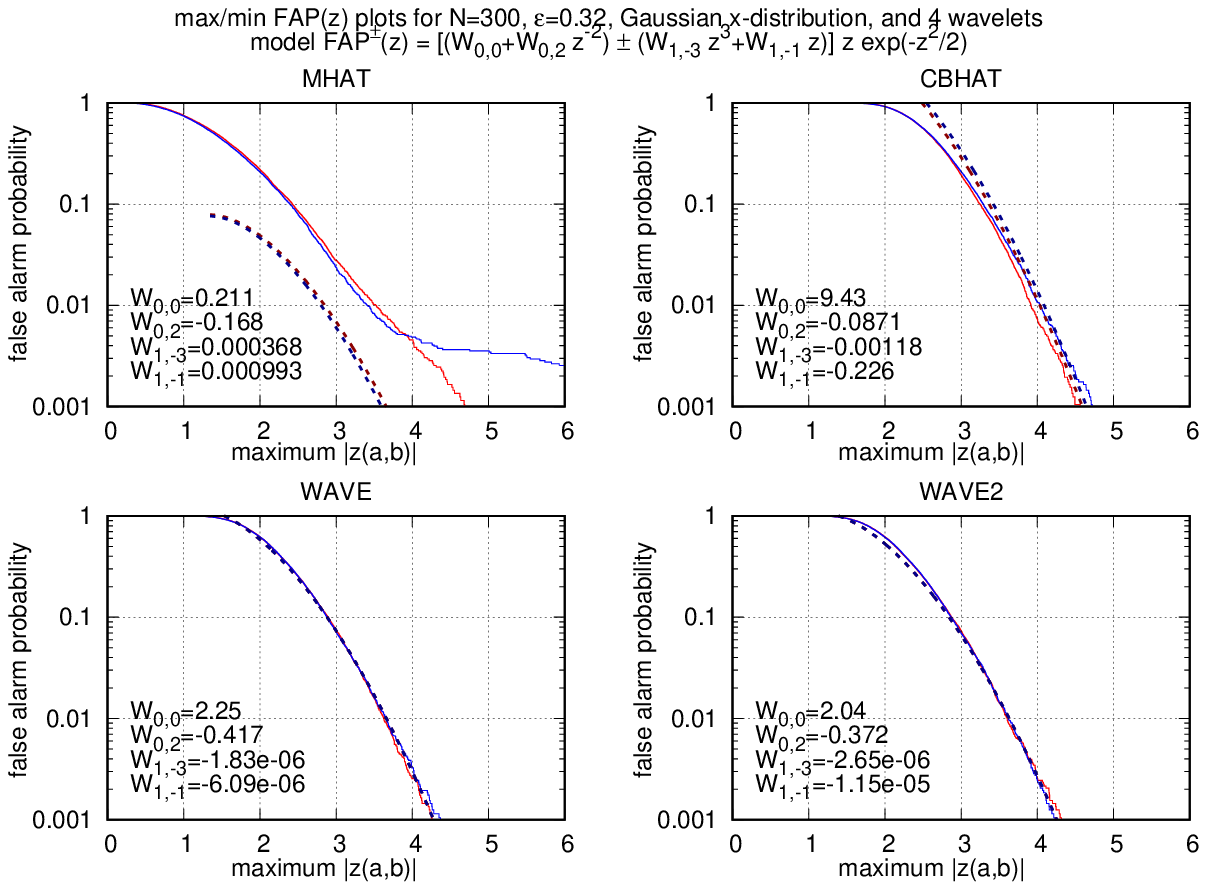}
\includegraphics[width=0.49\textwidth]{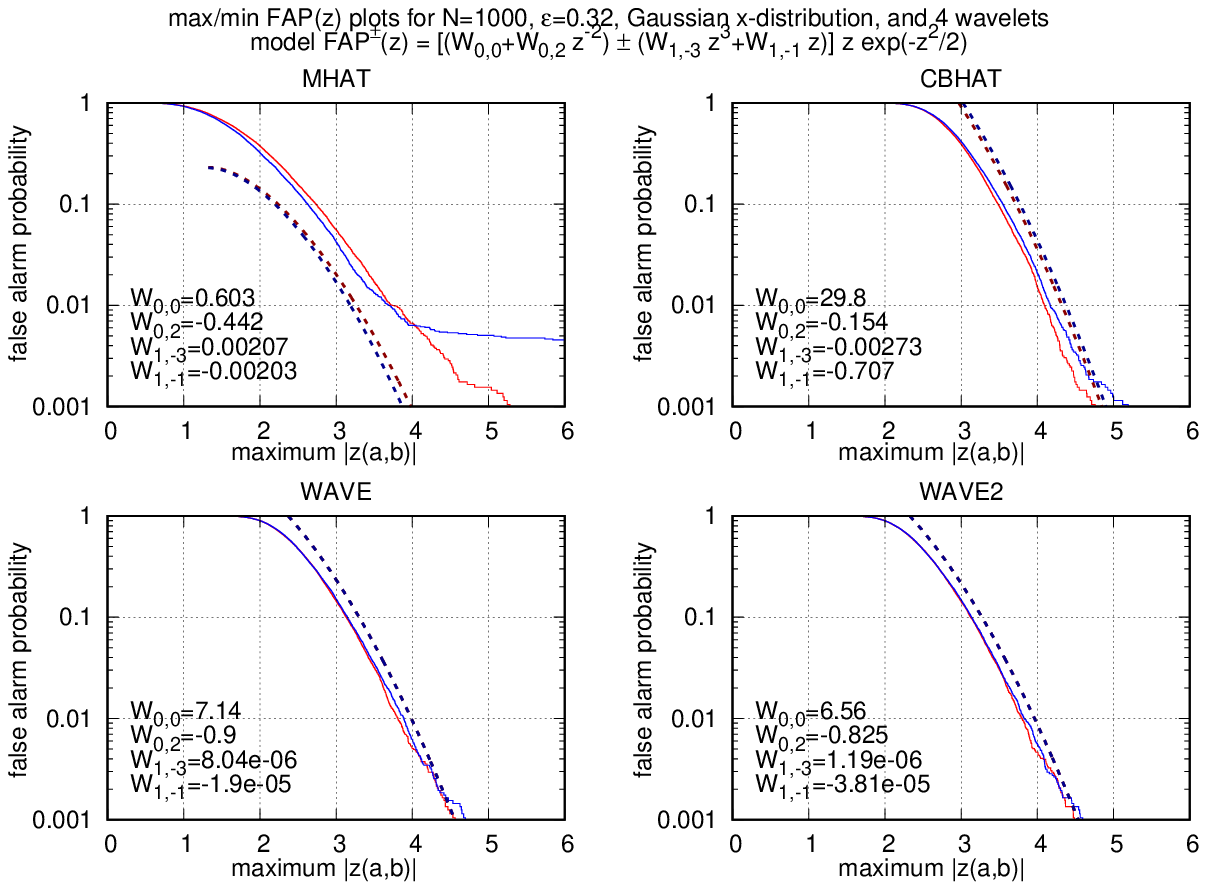}
\caption{Analytic approximations~(\ref{FAPgauss},\ref{FAPngauss}) of the $\FAP(z)$ law,
compared with Monte Carlo simulations. The simulations involved the normality
criterion~(\ref{nrmtest}) and thus all coefficients $W_{ij}$ are restricted to the
normality domain. Red-colored curves refer to positive extrema distributions (maxima of
$z(a,b)$, $\FAP^+$), blue-colored ones refer to the negative ones (minima of $z(a,b)$,
$\FAP^-$), though they appear mostly indistinguishable from each other. Theoretic
approximations are shown in thicker dashed lines.}
\label{fig_evdgauss}
\end{figure*}

We consider the accuracy of the $\FAP$ approximations in the Gaussian
case~(\ref{FAPgauss}), and partly~(\ref{FAPngauss}). Generating in the similar way $10^5$
simulated samples, for each trial we computed the maximum value $z_{\max}$ of $z(a,b)$. The
maximization was restricted to the normality domain, determined in accordance
with~(\ref{nrmtest}). The set of simulated $z_{\rm max}$ then was used to construct their
empirical distribution function and hence the simulated $\FAP$ curve. The latter can be
compared with the analytic approximations~(\ref{FAPgauss},\ref{FAPngauss}). This comparison
is shown in Fig.~\ref{fig_evdgauss}. We can see that the agreement is good, if $\FAP$ is
small (below $0.1$) and $W_{00}$ is not too small (above $1$). These are natural
restrictions of the Rice method used to approximate the $\FAP$. In particular, the MHAT
wavelet generates too small normality domain, implying that its $W_{00}$ is small as well,
and the analytic $\FAP$ approximation becomes poor. For other wavelets, we obtain good
agreement for the cases $N=300$ and $N=1000$. For $N=100$ (not shown), the quality of the
approximation degrades even for wavelets other than MHAT, because the normality domain
shrinks too much in any case. Therefore, our algorithm is applicable to samples that
contain a few hundred objects at least.

\begin{figure*}[!t]
\includegraphics[width=0.49\textwidth]{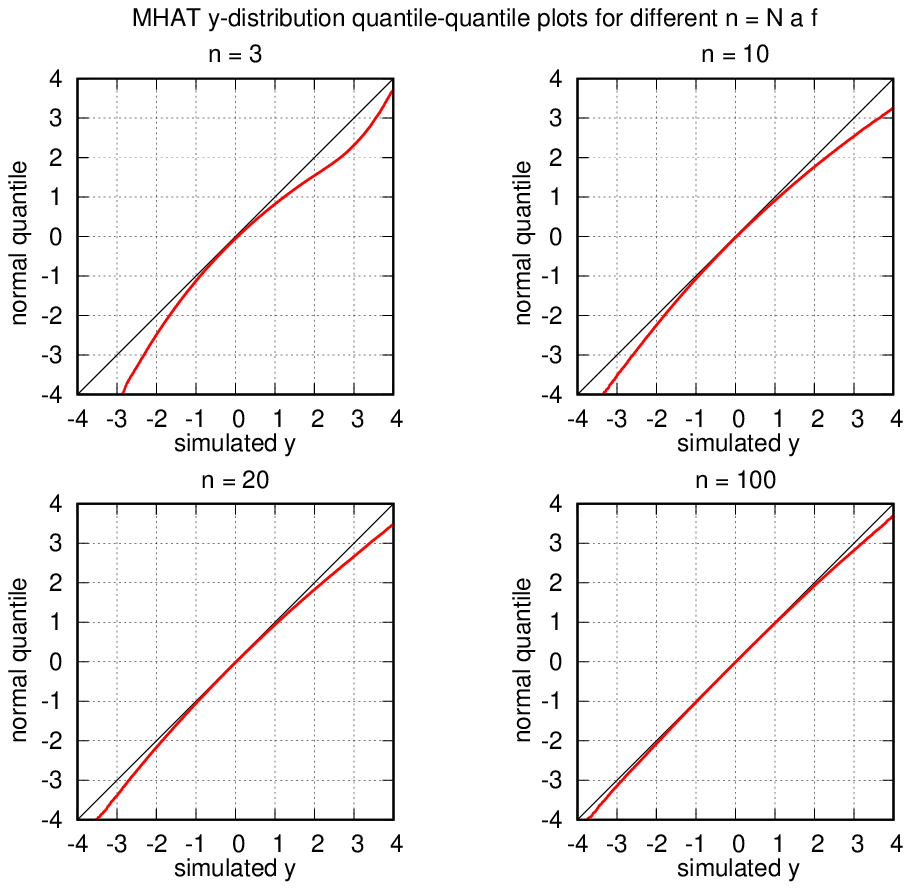}
\includegraphics[width=0.49\textwidth]{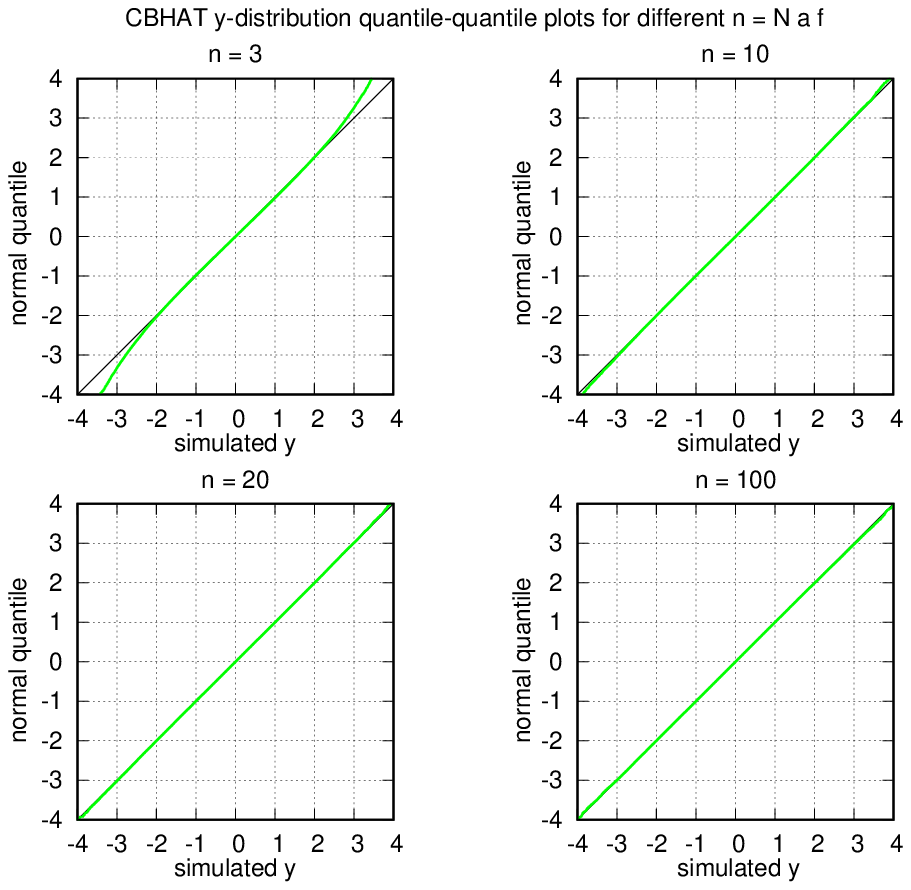}
\caption{Non-gaussianity effect inspired by the MHAT and CBHAT wavelets in the
$y$-distribution. These are normal quantile-quantile plots, in which the standard Gaussian
distribution should follow the main diagonal. Simulations assumed a flat $x$-distribution,
or $f(x)=\const$ within a wide range. The characteristic $n$ here is the math. expectation
of~(\ref{num}), that is the average number of samples $x_i$ falling in the wavelet
localization domain.}
\label{fig_yqq}
\end{figure*}

The effect of the wavelet normality is demonstrated in Fig.~\ref{fig_yqq}. We show the
$y$-distribution generated by the MHAT or CBHAT wavelet, depending on the characteristic
$n$ from~(\ref{num}). In the first case, this distribution is clearly non-Gaussian, and
this remains noticable even for $n$ as large as $100$. The CBHAT wavelet demonstrates much
better behaviour: the non-Gaussian deviations become invisible to an eye for $n=10$
already. This confirms that the non-Gaussianity in CBHAT was indeed suppressed a lot.

Finally, we apply our technique to a real-world data set. We considered the exoplanetary
candidates from the \emph{Extrasolar planets catalog} at \texttt{www.exoplanet.eu}. The
sample was formed from $N=695$ candidates detected by the radial-velocity technique. Using
our wavelet method, we analysed the distribution of orbital periods $P$ of these candidates
(more precisely, we considered $x = \log P$). We assumed the simplest initial model
$Y_0\equiv 0$.

We do not plan to analyse this distribution here in depth, and we currently avoid any
physical interpretation of the results (this is left for the more thorough work in future).
Our aim is only to further test the wavelet analysis method and demonstrate how it may work
with real data. So we did not undertake any attempts to `clean' this sample to make it more
homogeneous or less affected by observational biases. We also assume that thanks to the
third Kepler law, the orbital periods are equivalent, in average, to the distance from the
star (orbital semimajor axis), because majority of the stars involved in the planet search
programmes are similar to the Sun and have roughly the same mass of $1$ Solar mass. So in
the statistical sense a `short-period' planet is roughly the same as a `close-in' planet.

\begin{figure*}[!t]
\includegraphics[width=0.49\textwidth]{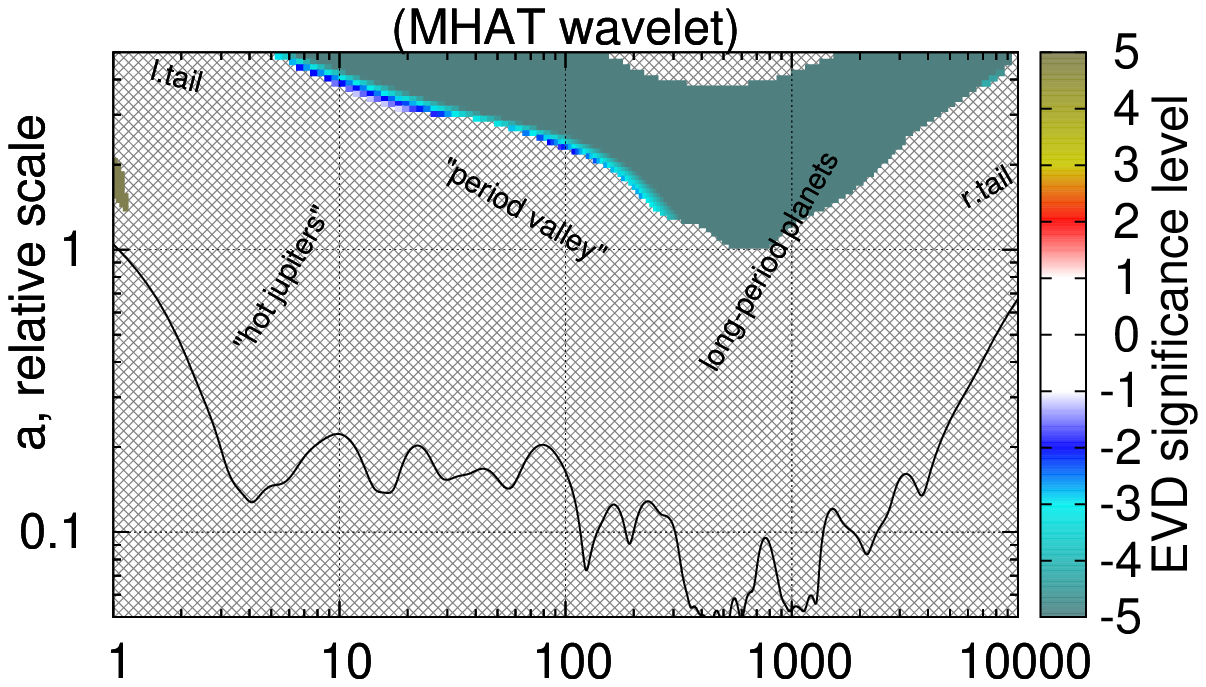}
\includegraphics[width=0.49\textwidth]{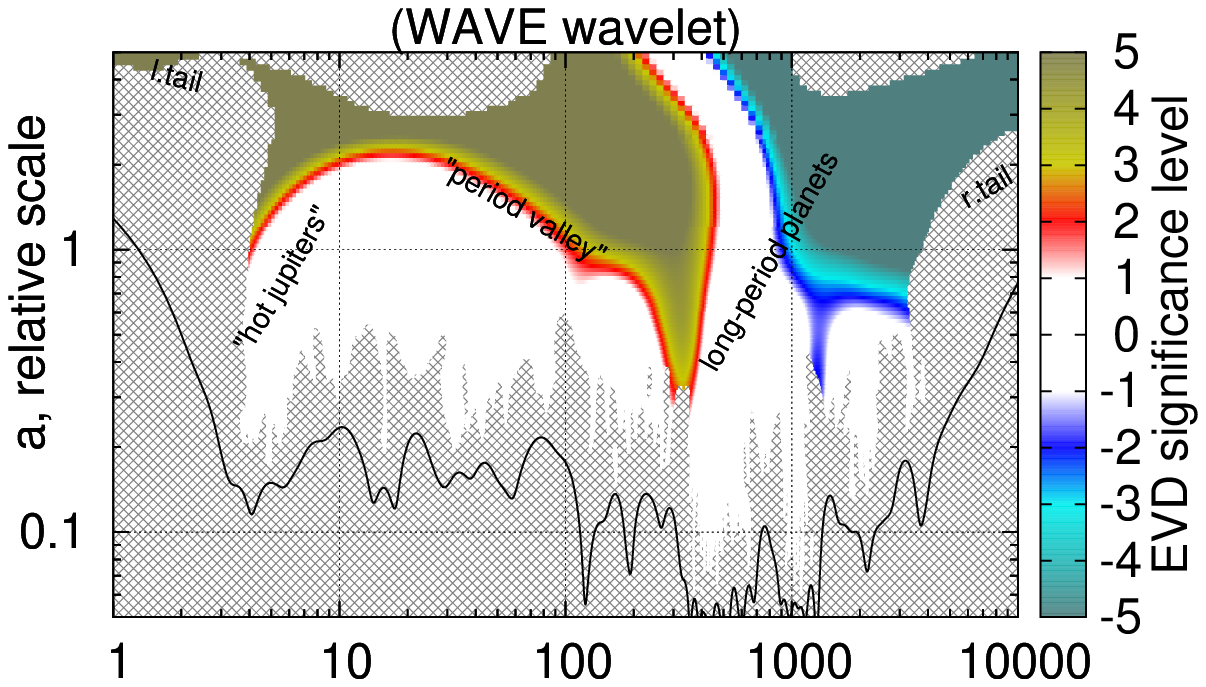}\\
\includegraphics[width=0.49\textwidth]{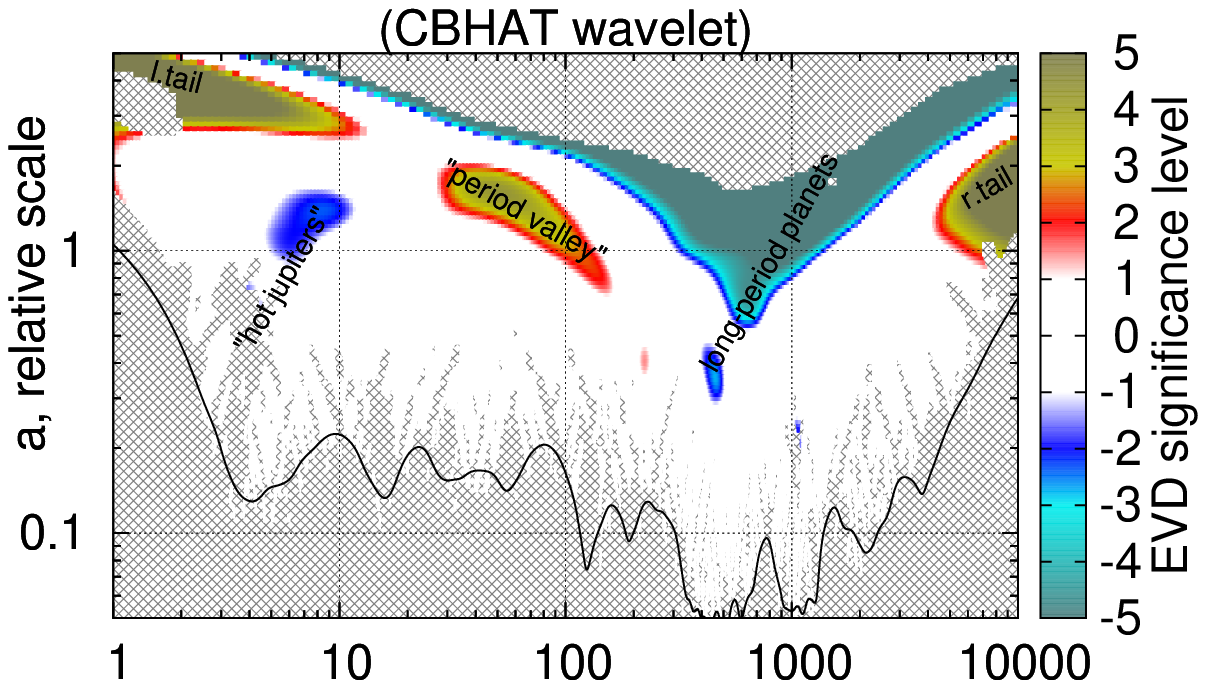}
\includegraphics[width=0.49\textwidth]{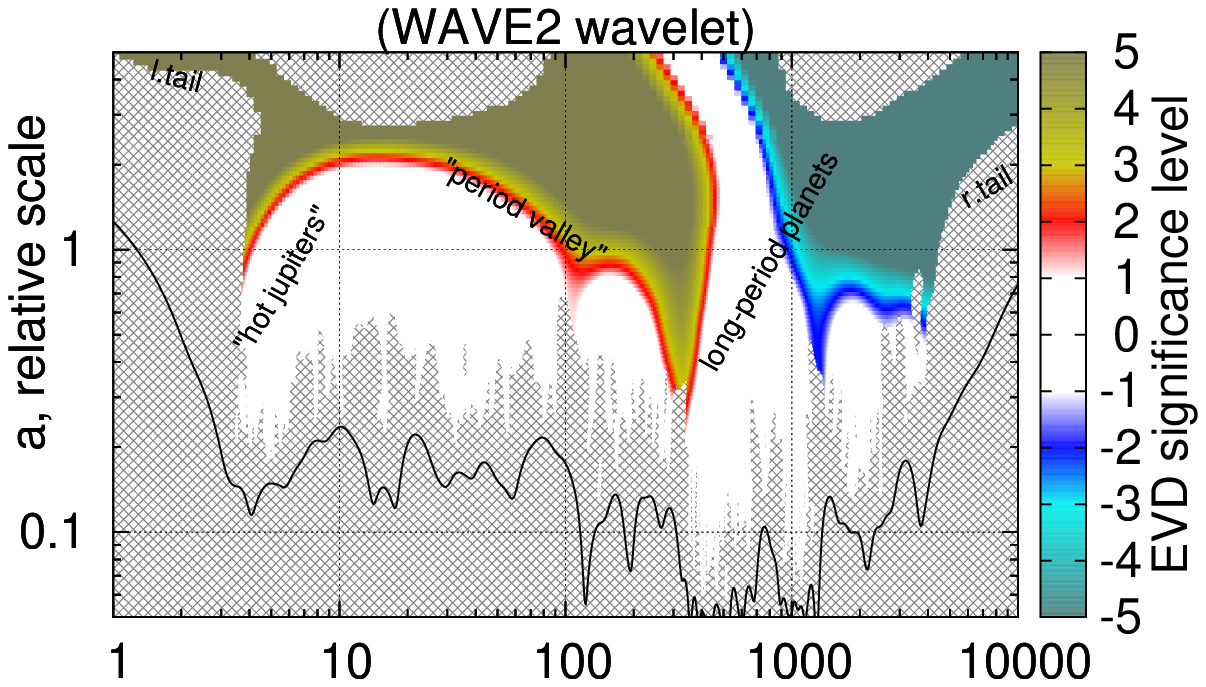}\\
\includegraphics[width=0.49\textwidth]{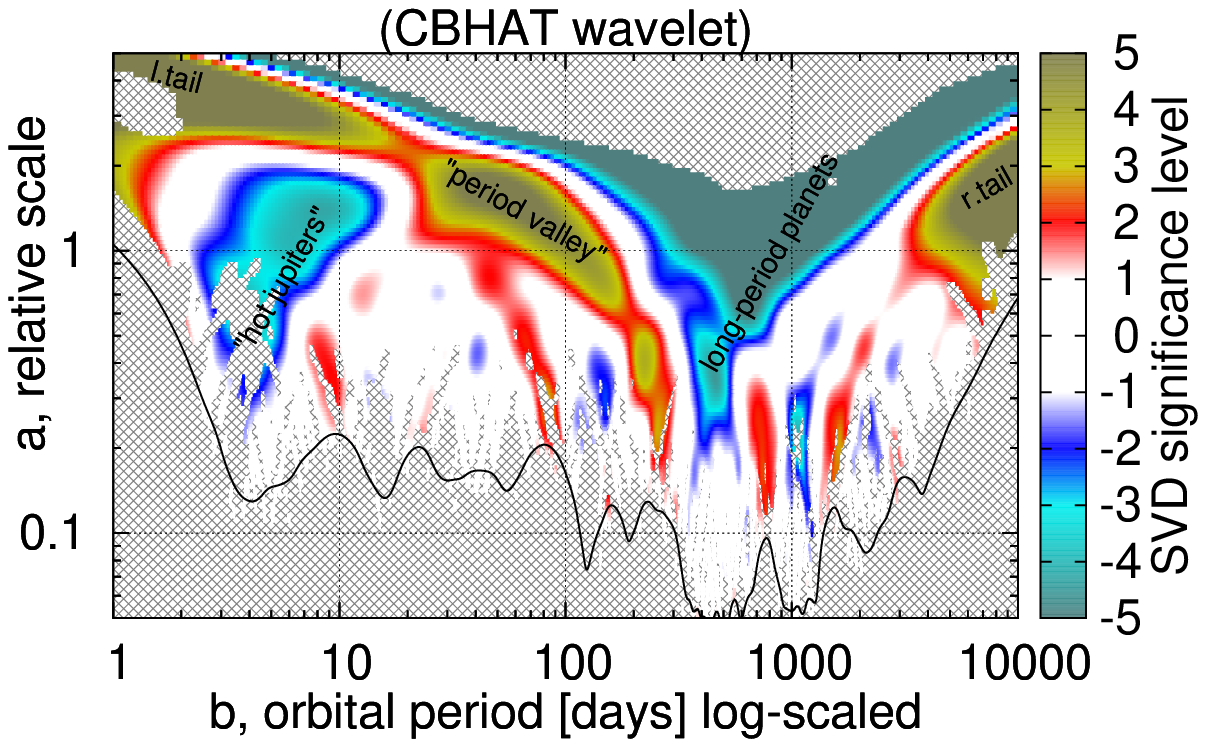}
\includegraphics[width=0.49\textwidth]{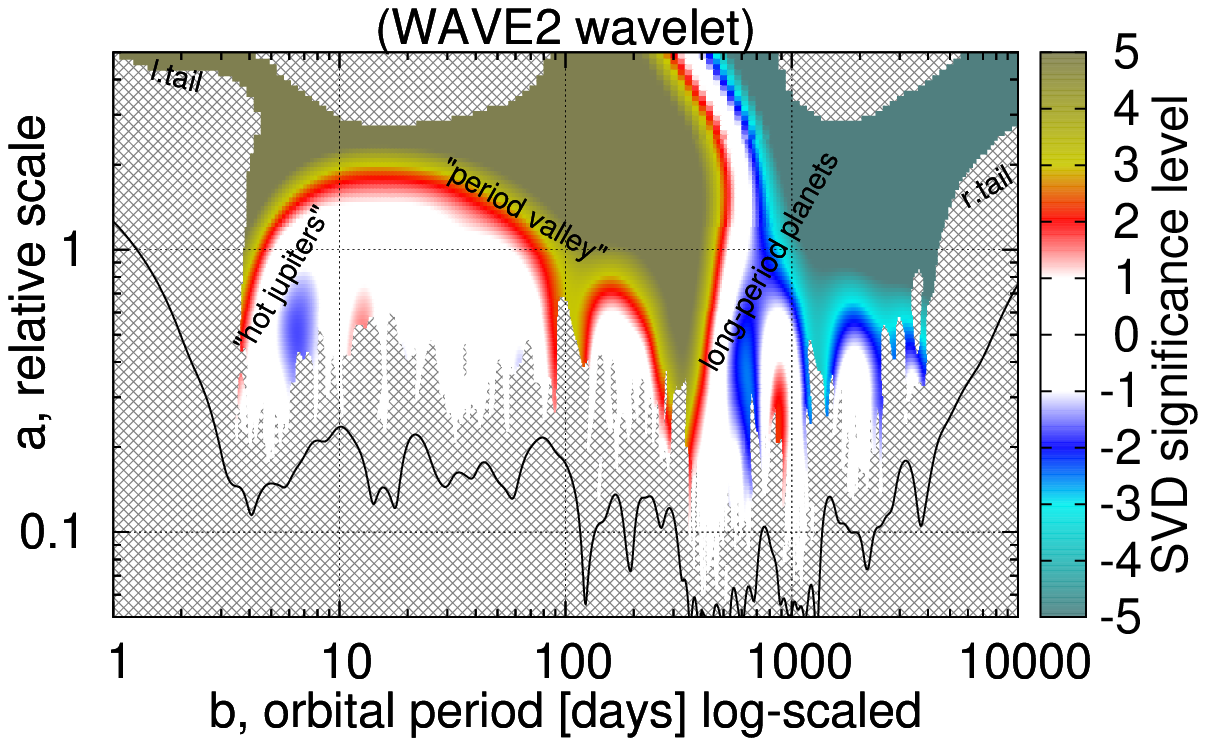}
\caption{Wavelet transforms of the orbital periods distribution for $N=695$ exoplanetary
candidates. The colors map the significance of $z(a,b)$, taking the sign into account
(red-yellow for $z>0$, blue-cyan for $z<0$). Two bottom panels show the `SVD
significance'$=z$. The others show the `EVD significance', being the normal quantile of the
$\FAP$ estimate~(\ref{dsFAPgauss}). The hashed regions are where the normality
test~(\ref{nrmtest}) was failed. The black curves near the bottom of each graph mark the
condition $n=10$, with $n$ given in~(\ref{num}).}
\label{fig_periods_wt}
\end{figure*}

Our analysis algorithm provides two types of the output: (i) the normalized wavelet
transform~(\ref{zdef}), accompanied by the significance estimate~(\ref{dsFAPgauss}), and
(ii) the reconstructed p.d.f. obtained iteratively as described in Sect.~\ref{sec_alg}
below. We insist that the wavelet transform should be treated as the primary source of
information, because it allows to clearly separate different patterns from each other, and
also provides a direct estimate of their statistical significance. The reconstructed p.d.f.
is merely a nice and intuitive representation of this wavelet map. This reconstruction is
still not completely free from artifacts, and it was not designed to give any significance
measures.

Various wavelet transforms of this sample are plotted in Fig.~\ref{fig_periods_wt}. Based
on these graphs, we may highlight the following:
\begin{enumerate}
\item The MHAT wavelet is practically useless in this analysis, because it generates too
large non-Gaussianity.

\item Other wavelets appear quite useful and easily identify several well-known structures
in the exoplanetary period distribution \citep{CummingStat}: the primary maximum containing
long-period exoplanets ($P\gtrsim 1$~yr, very large significance), the family of `hot
jupiters' at short periods ($P\sim 3-10$~d, approx. two-sigma significance), and the
`period valley' between them (above the three-sigma significance). The small-scale side
spike near the long-period group (in CBHAT plot, about three-sigma significance) indicates
that there is a relatively sharp transition between them and the period valley.

\item The effect of the `domain penality' on the significance estimates is dramatic. We
obtain considerably more diverse set of structures, if the significance is determined from
the single-value $\FAP$ distribution, like in \citet{Skuljan99}. However, these increased
significance estimates would correspond to an impractical condition that the scale and
position of a given structure was known \emph{a priori}. Most of the structures appearing
significant in the SVD mode, become just usual noisy deviations in the EVD mode. In
practice the SVD significance is usually inadequate, because we basically \emph{estimate}
patterns locations by spotting the maximum deviations over the plot.

\item Our method could in principle detect in this sample details as small as $a\sim 0.05$,
based on the minimum scale allowed by the normality criterion, though small-scale
structures can be detected only if they have enough amplitude.

\item The graphs constructed with the use of an odd or an even wavelet appear highly
complementary. One allows to better understand another.
\end{enumerate}

\begin{figure*}[!t]
\includegraphics[width=0.49\textwidth]{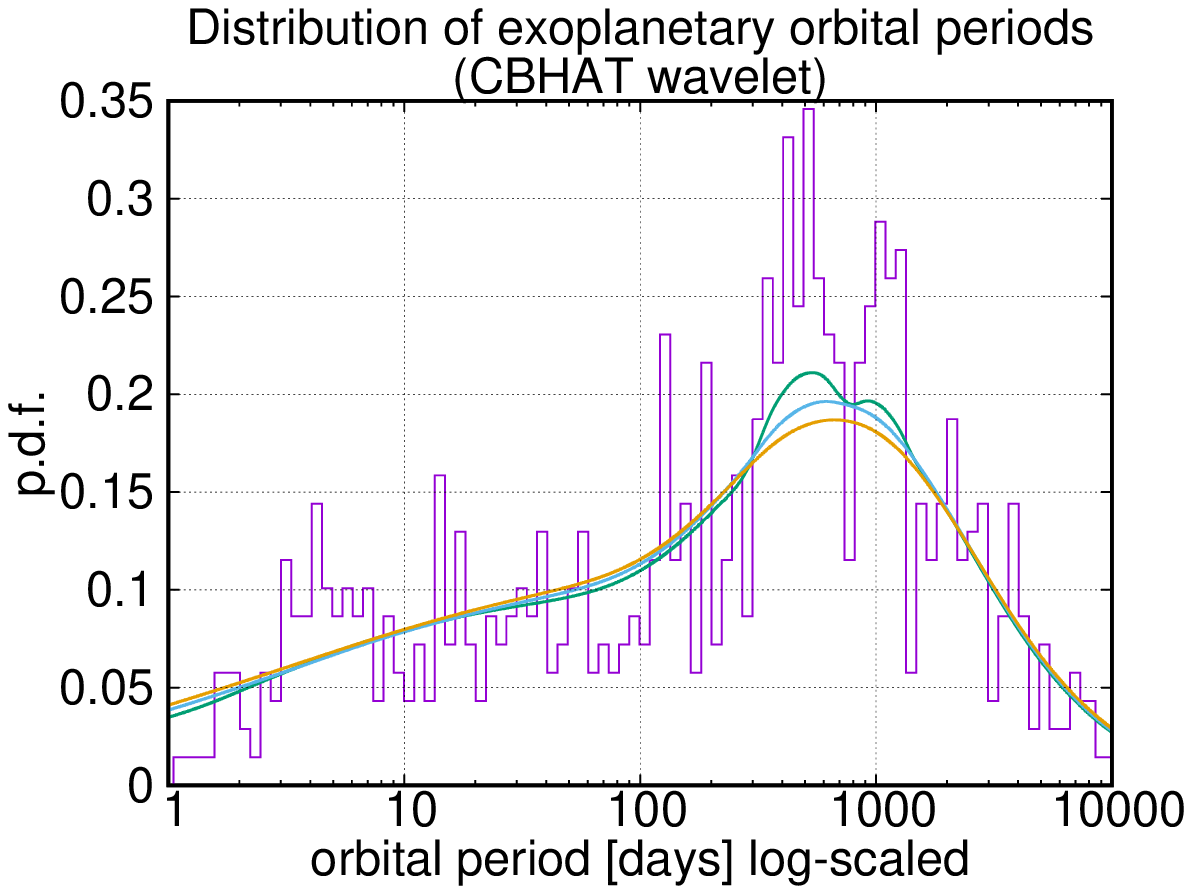}
\includegraphics[width=0.49\textwidth]{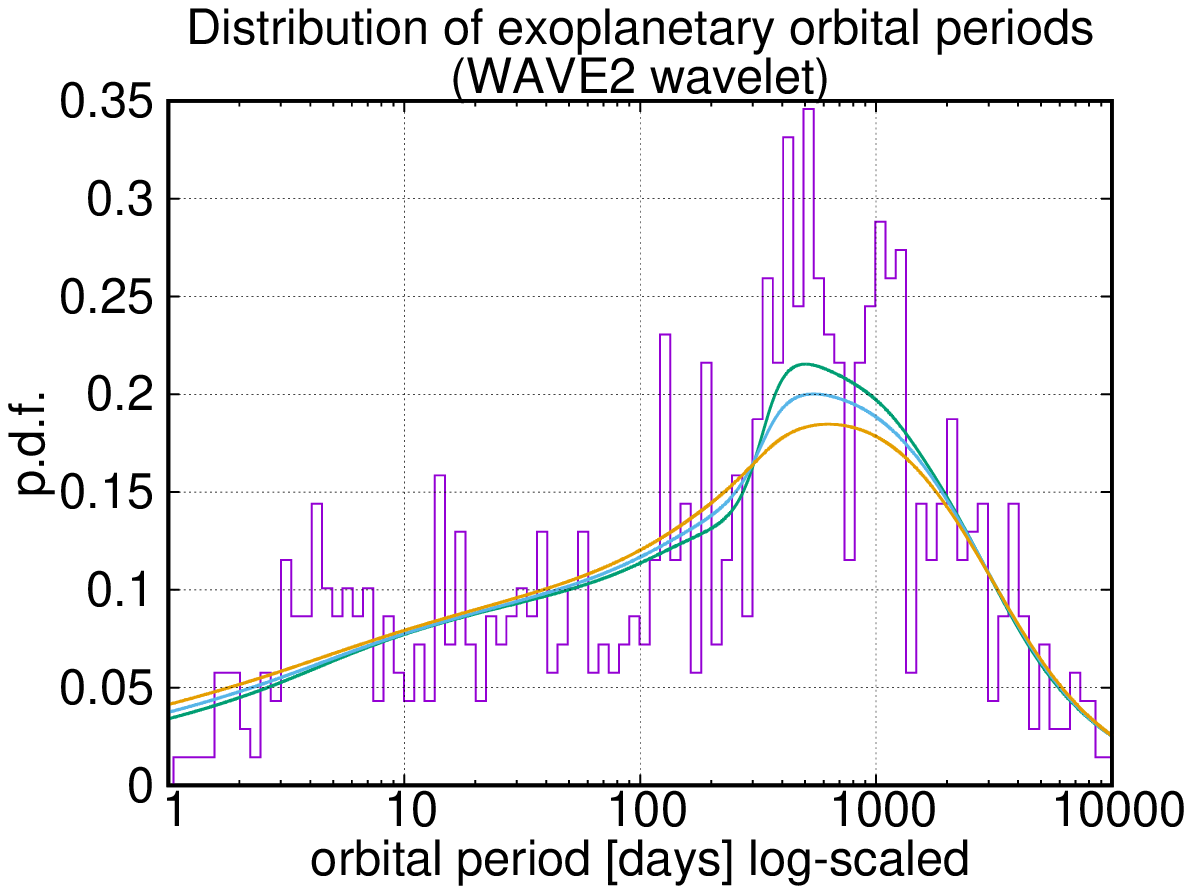}
\caption{Reconstructed p.d.f. of the exoplanetary orbital periods distribution, based on
the same sample as in Fig.~\ref{fig_periods_wt}. The soft thresholding method was used
here. Different curves refer to one-sigma ($\FAP=0.31$), two-sigma ($\FAP=0.05$), and
three-sigma ($\FAP=0.0027$) signficance tolerances. A histogram of the sample is also
plotted.}
\label{fig_periods_rec}
\end{figure*}

The reconstructed p.d.f. of this distribution is plotted in Fig.~\ref{fig_periods_rec} for
WAVE2 and CBHAT wavelets and three different noise threshold levels, corresponding to the
$1,2,3$-sigma significance. We can see that the `hot jupiters' density bump, as well as the
`period valley' depression, both appear rather shallow in the absolute measure.
Nevertheless, their significance in the SWT is high enough. An interiguing detail is a
possible bimodality of the primary maximum that appears at the $1$-sigma reconstruction
with the CBHAT wavelet. However, the wavelet transform did not contain any hint of a sharp
local minima at this position ($P\sim 700$~d), even at the $1$-sigma significance level.
This means that this local minima may be a minor artifact of the reconstruction. It is the
wavelet transform that provides us with the significance information, so we must wait until
this putative local minimum reveals itself in $z(a,b)$. Nevertheless, this period range
definitely contains subtle details that need further interpretation, in particular there is
a rather sharp border with the `period valley'. The advantage of our wavelet technique is
that it allows us to discuss such details in a much more rigorous and objective manner than
e.g. the traditional histogram.

A detailed analysis of a wide variety of exoplanetary distributions is given in
\citep{Baluev18b}.

\section{Summary of the algorithm}
\label{sec_alg}
In this section, we provide a brief summary of the method with quick references to the main
formulae. A software project implementing the entire pipeline of the algorithm is hosted at
the \texttt{SourceForge.net} service under the title {\sc
WaveletStat}.\footnote{\texttt{http://sourceforge.net/projects/waveletstat/}}

The algorithm infers the use of one of the optimized wavelets, WAVE2~(\ref{WAVE2}) or
CBHAT~(\ref{CBHAT}), normalized according to either~(\ref{oddnrm}) or~(\ref{evennrm}). The
analysis requires to solve four major sub-tasks described below. Note that algorithmically
they may overlap with each other.

\subsection{Computation of the SWT}
\label{alg_samplewt}
Use formulae~(\ref{wavest},\ref{wavvarest},\ref{zdef}) to compute $\widetilde Y(a,b)$ and
$z(a,b)$ on some dense enough grid in the $(a,b)$ plane. The grid has to be non-uniform: it
should increase its $b$-axis density to smaller $a$ (as $\Delta b \propto a$), and the
$a$-axis should be logarithmic at small $a$, in order to have all local maxima of the CWT
sampled uniformly. A good choice is a uniform grid in $\log(1+a_0/a)$, where $a_0$ is
proportional to the sample variance.

\subsection{Determination of the normality domain}
\label{alg_gaussdmn}
The normality domain in the $(a,b)$ plane is determined based on the
criterion~(\ref{nrmtest}), in which the quantities $q_{ij}$ are defined in \ref{sec_edge}:
\begin{enumerate}
\item Expressions~(\ref{v}) define the vector $\bmath v$.
\item An estimation $\tilde{\bmath v}$ is constructed by substituting the covariance
estimates~(\ref{covest}) and their necessary analogues, including $Y\simeq \widetilde Y$.
\item The elements of $\tilde{\bmath v}$ are arranged in products and then are
sample-averaged according to~(\ref{kumest}), yielding various momenta estimates.
\item Finally, $q_{ij}$ are constructed from these momenta using formulae~(\ref{qij}) and
those given in the attached MAPLE worksheet.
\end{enumerate}
It is also necessary to set in~(\ref{nrmtest}) two error control parameters $z_*$ and
$\varepsilon$. Values $z_*^2=10$ and $\epsilon^2=0.1$ appear practically reasonable.

\subsection{Applying a $\FAP$ threshold to clean the noise}
\label{alg_fapclr}
First we need to compute $W_{00}$ in~(\ref{W00}), substituting $\det\mathbmss G$
from~(\ref{qij}) and other quantities already defined. The integration in~(\ref{W00})
should be made within the normality domain. Optionally, a correction term $W_{02}$ may be
computed, based on~(\ref{Wij}) with $q_{02}$ expressed in the attached MAPLE worksheet.
After that, apply the asymptotic estimate~(\ref{dsFAPgauss}) to compute the $\FAP$ for each
points inside the normality domain. Threshold the points given some small critical
$\FAP_{\rm thr}$: domains with $\FAP<\FAP_{\rm thr}$ correspond to significant features in
the CWT, while everything else should be attributed to the noise. This can be made for
different $\FAP_{\rm thr}$.

\subsection{Reconstruction of the p.d.f. $f(x)$ from the cleaned wavelet transform}
Apply the hard or soft noise thresholding to $\widetilde Y$ to obtain $Y_{\rm thr}$. Then
substitute it in the inversion formula~(\ref{wavinv}), using the optimized reconstruction
kernel $\gamma$ defined in~(\ref{gamma}), and the constants from~(\ref{Cpg}). After this,
we obtain a better approximation than $Y_0$ was, but it is not guaranteed that no
significant patterns remain in the residual wavelet transform. We should update $Y_0$ with
this new model and return back to the noise thresholding step~\ref{alg_fapclr}, iterating
until all significant patterns (in the normality domain) are cleared away from the
residual. The SWT from the step~\ref{alg_samplewt} and the Gaussian domain
from~\ref{alg_gaussdmn} do not change in these iterations.

\section{Conclusions and discussion}
Our main results that allowed to significantly improve available methods of the p.d.f.
wavelet analysis, are:
\begin{enumerate}
\item Asymptotic estimation of the p-value significance, obtained with an adequate
treatment of the `domain penalty' effect. This approximation is practically accurate and
entirely analytic, thus removing the need of Monte Carlo simulations.
\item Construction of an objective criterion to determine the applicability domain of the
method in the shift-scale plane, based on the Gaussianity requirement.
\item Derivation of optimal wavelets and optimal reconstruction kernels that allowed to
improve the S/N ratio and to reduce the non-Gaussian deviations, thus expanding the
applicability domain in the shift-scale plane. We also showed that the MHAT wavelet is
practically useless in this task because of the large non-Gaussianity it generates.
\end{enumerate}
The simulations revealed that the method can be used on samples containing a few hundred of
objects at least, because otherwise the normality domain in the $(a,b)$ plane shrinks to
much, rendering the analysis unreliable. Another limitation of our algorithm is that it can
process only 1D distributions. It cannot analyse 2D distributions that were the main goal
in e.g. \citep{Skuljan99}, or distributions of higher dimension.

Generalization of the technique to two dimensions and more remains a task for future, and
the above theory represents the necessary basis for future improvements. The design of the
method would probably allow us to incorporate a neat reduction of various statistical
distortions, e.g. selection biases. In the current form our method can be already applied
to exoplanetary distributions, including candidates discovered by the Kepler spacecraft, as
well as to study the Milky Way stellar population.

\section*{Acknowledgements}
This work was supported by the Russian Foundation for Basic Research grant 17-02-00542~A
and by the Presidium of Russian Academy of Sciences programme P-28, subprogramme ``The
space: investigating fundamental processes and their interrelations''.

\appendix

\section{Edgeworth-type expansion for the false alarm probability}
\label{sec_edge}
Let $\nu=1/N$, and $\mu=\sqrt\nu$. Then the classic Edgewort series is a power series in
$\mu$ for the distribution function of a sum of idependent identically distributed
variables. This is, it is a power series of the $y$-distribution in our case. However, we
need to deal with multivariate distributions involving also derivatives of $y$, and
construct a multivariate version of the Edgeworth series. So let us define an auxiliary
vector:
\begin{align}
\bmath v &= \left( y_{\rm nrm}, \frac{\partial y_{\rm nrm}}{\partial b}, \frac{\partial y_{\rm nrm}}{\partial a}, \frac{\partial^2 y_{\rm nrm}}{\partial b^2}, \frac{\partial^2 y_{\rm nrm}}{\partial b \partial a}, \frac{\partial^2 y_{\rm nrm}}{\partial a^2} \right), \nonumber\\
y_{\rm nrm} &= \frac{y-Y}{\sqrt{\disp y}}, \qquad
y_{\rm nrm}' = \frac{y' - Y'}{\sqrt{\disp y}} - y_{\rm nrm} \frac{\cov(y,y')}{\disp y}, \nonumber\\
y_{\rm nrm}'' &= \frac{y''-Y''}{\sqrt{\disp y}} - y_{\rm nrm} \frac{\cov(y,y'')+\var y'}{\disp y} - \nonumber\\
&\quad - \frac{y_{\rm nrm}' \otimes \cov(y,y') + \cov(y,y') \otimes y_{\rm nrm}'}{\disp y},
\label{v}
\end{align}
where $\otimes$ stands for the dyadic product of vectors.

Consider the multivariate cumulant-generating function (c.g.f.) of the vector~(\ref{v}):
\begin{equation}
K_{\bmath v}(\bmath t) = \log(\expect e^{\bmath t \cdot \bmath v}) =
 \frac{1}{2!} {\bmath l}_2 \cdot \bmath t^2 + \frac{1}{3!} {\bmath l}_3 \cdot \bmath t^3 + \frac{1}{4!} {\bmath l}_4 \cdot \bmath t^4 + \ldots
\label{KGF}
\end{equation}
Here, $\bmath t$ is a vector of dimension $d=6$ (same as $\dim\bmath v$), and ${\bmath
l}_n$ are symmetric tensors of degree $n$ and dimension $d$, starting from a $d\times d$
matrix ${\bmath l}_2$. Each ${\bmath l}_n$ contains $C_{n+d-1}^{d-1}$ of independent scalar
elements, representing various cumulants of $\bmath v$. In particualar, ${\bmath l}_2$ is
the covariance matrix $\var \bmath v$. Since $\expect \bmath v=0$ by definition, we have
${\bmath l}_1=0$. These cumulants are our input parameters; they are estimated by various
$L_{ijk}$ in~(\ref{kumest}).

In accordance with~(\ref{Rice}) we need to approximate the distribution of the vector
$(z,z',z'')$, rather than $\bmath v=(y,y',y'')$. Therefore, we need to transform one c.g.f.
to another. This transformation has to be done in a few intermediate steps layed out below.

The $z$ definition~(\ref{zdef}) and its derivatives contain several sample momenta that
involve various products of $y$, $y'$, and $y''$. Therefore, at first we should depart from
these quantities to an extended set containing various their powers and products. We do not
know yet, what exactly products we will need, so we just denote this new vector as ${\bmath
\omega}$. Its c.g.f. can be defined analogously to~(\ref{KGF}). Now let us consider the
moment-generating function (m.g.f.) of $\bmath v$:
\begin{align}
M_{\bmath v}(\bmath t) &= \exp[K_{\bmath v}(\bmath t)] = \expect e^{\bmath t \cdot \bmath v} = \nonumber\\
&= 1 + \frac{1}{2!} {\bmath m}_2 \cdot \bmath t^2 + \frac{1}{3!} {\bmath m}_3 \cdot \bmath t^3 + \frac{1}{4!} {\bmath m}_4 \cdot \bmath t^4 + \ldots
\label{MGF}
\end{align}
with a similar series for $M_{\bmath \omega}(\bmath t)$. Each coefficient in~(\ref{MGF})
represents a tensor containing various momenta of $\bmath v$. The expansion~(\ref{MGF}) can
be obtained by exponentiating~(\ref{KGF}) with a subsequent development into Taylor series.
This allows to express all ${\bmath m}_n$ via ${\bmath l}_n$. However, by construction of
$\bmath\omega$, each its momentum is equal to some momentum of $\bmath v$. That is, there
is a one-to-one mapping between the coefficients of the m.g.f. series~(\ref{MGF}) for
$\bmath v$ and of a similar series for ${\bmath \omega}$. This property allows us to
express the m.g.f. $M_{\bmath \omega}(\bmath t)$ via the cumulants ${\bmath l}_n$. After
that, we only need to expand $\log M_{\bmath \omega}$ to obtain a series for the c.g.f.
$K_{\bmath \omega}$.

Now we must make a transition from $\bmath\omega$ to the sample average $\bmath\Omega =
\langle\bmath\omega\rangle$, because $z$ is defined via sample averages. At this point we may
use the primary property of the c.g.f.: whenever independent random quantities are being
added, their c.g.f.'s are plainly summed up, as well as the cumulants. The c.g.f. of a sum
of identically distributed variables is just a multiple of the original c.g.f. In our case,
we also have a division by $N$ in the averaging operation $\langle *\rangle$, so we obtain
\begin{equation}
K_{\bmath\Omega}(\bmath t) = N K_{\bmath\omega}(\bmath t/N).
\label{KGFavg}
\end{equation}
We represented $K_{\bmath\omega}$ in the form of a Taylor series, so by rearranging terms
in the right hand side of~(\ref{KGFavg}) we can obtain a power series with respect to the
minor parameter $\nu$. Note that contrary to~(\ref{KGF}), $K_{\bmath\omega}$ \emph{does
have} a non-zero first-order term in the expansion, because some averages of $\bmath\omega$
do not vanish (for example those that involve even powers of $v_i$). This implies that the
expansion in~(\ref{KGFavg}) starts from the power $\nu^0$, and this term must be linear in
$\bmath t$:
\begin{equation}
K_{\bmath\Omega}(\bmath t) = \bmath\Omega_c \cdot \bmath t + \mathcal O(\nu)
\end{equation}
The coefficients $\bmath\Omega_c$ represent asymptotic limit of the average
$\expect\bmath\Omega$ for $N\to\infty$, whereas all other cumulants, including the
variances, tend to zero. That is, $\bmath\Omega\to\bmath\Omega_c$ whenever $N\to\infty$, as
expected for sample averages.

According to~(\ref{zdef}), the $z$ statistic depends on just the two sample momenta,
$\Omega_0 = \langle y_{\rm nrm} \rangle$ and $\Omega_{00} = \langle y_{\rm nrm}^2
\rangle$. When taking the derivatives $z'$ and $z''$ we obtain more sample momenta of
higher degrees, involving derivatives of $y_{\rm nrm}$. At this point we finally can
determine, which products we actually needed to include in the $\bmath\omega$ vector: they
should correspond to the sample momenta $\bmath\Omega$ that appeared in $z$, $z'$, and
$z''$. Algorithmically, this part has to be done before computing $K_{\bmath\omega}$, of
course.

Obviously, $z$ and its derivatives are non-linear functions of $\bmath\Omega$. To
approximate their momenta or cumulants, we follow \citet{Martins10} and employ the
so-called delta method. First, we expand the vector $\bmath z=(z,z',z'')$ into a power
series about $\bmath\Omega = \bmath\Omega_c$:
\begin{equation}
\bmath z(\bmath\Omega) = \bmath z(\bmath\Omega_c) + \bmath z'(\bmath\Omega_c) \cdot \Delta\bmath\Omega + \frac{1}{2}\bmath z''(\bmath\Omega_c) \cdot \Delta\bmath\Omega^2 + \ldots
\label{zpwr}
\end{equation}
Here, $\Delta\bmath\Omega = \bmath\Omega - \bmath\Omega_c$ is a small quantity with the
magnitude $\mathcal O(\mu)$. Now, let us substitute the series~(\ref{zpwr}) in the
definition of the $\bmath z$-vector m.g.f.:
\begin{equation}
M_{\bmath z}(\bmath t) = \expect e^{\bmath t \cdot \bmath z},
\label{zMGF}
\end{equation}
and develop the exponent into a power series with respect to $\Delta\bmath\Omega$. After
taking the mathematical expectation operator, the coefficients of this series become equal
to various momenta of $\Delta\bmath\Omega$. On the other hand, these momenta can be
determined as coefficients in an analogous series, obtained by expanding the m.g.f.
\begin{align}
M_{\Delta\bmath\Omega}(\bmath t) &= \exp[K_{\Delta\bmath\Omega}(\bmath t)], \nonumber\\
K_{\Delta\bmath\Omega}(\bmath t) &= K_{\bmath\Omega}(\bmath t) - \bmath\Omega_c \cdot \bmath t = \mathcal O(\nu).
\label{KMD}
\end{align}
It is easier to expand~(\ref{KMD}) with respect to a scalar minor parameter $\nu$ first. In
this decomposition, the coefficient collected near each $\nu^n$ is a polynomial in $\bmath
t$. We should rearrange the series then, in order to collect the coefficients near $\bmath
t$-powers. Each such a coefficient will represent a power-series in $\nu$, all truncated at
the same $\nu$-power. We need to substitute these $\bmath t$-power coefficients into
appropriate positions of the series for~(\ref{zMGF}). Note that $\bmath z$ also depends
explicitly on $\nu$ via $\sqrt{N-1}\sim\mu$, so the decomposition for~(\ref{zMGF}) should
be made simultaneously with respect to $\mu$ rather than $\nu$, and $\Delta\bmath\Omega\sim
\mu$. At this point we must take a special care about the truncation of the $\mu$-series
that must occure at the same $\mu$-order in each term being added. This procedure results
in the m.g.f. $M_{\bmath z}(\bmath t)$ being expressed via the original cumulants $l_n$,
and up to a certain $\mu$-order $\mu^n = N^{-n/2}$.

Now, by making a simple substitution
\begin{equation}
\chi_{\bmath z}(\bmath t) = M_{\bmath z}(i\bmath t)
\label{chiz}
\end{equation}
we obtain the characteristic function of the vector $\bmath z$, which is a Fourier
transform of the multivariate p.d.f. $p_{zz'z''}$. Therefore, to obtain the final result we
may apply inverse Fourier to the expansion of~(\ref{chiz}), and then substitute the result
to~(\ref{Rice}). However, it appears easier to perform the integration~(\ref{Rice})
directly in the Fourier space, making use of the Parseval identity:
\begin{equation}
\int\limits_{-\infty}^{+\infty} f(s) g^*(s) ds = \frac{1}{2\pi} \int\limits_{-\infty}^{+\infty} \hat f(t) \hat g^*(t) dt
\end{equation}
In our case one of the functions in the right hand side is $\chi_{\bmath z}(\bmath t)$, and
the other one should be the Fourier transform of everything that should remain inside the
integrals in~(\ref{Rice}), if we remove $p_{zz'z''}$:
\begin{equation}
(z''_{aa} z''_{bb} - {z''_{ab}}^2) \delta(z'_a) \delta(z'_b) U(z-z_{\rm thr}),
\label{mlt}
\end{equation}
where $\delta(x)$ is the Dirac delta function, and $U(x)$ is the Heaviside function. The
Fourier transform of~(\ref{mlt}) with respect to $\bmath z =
(z,z'_a,z'_b,z''_{aa},z''_{ab},z''_{bb})$ reads:
\begin{align}
(2\pi)^3 \left(\pi \delta(t_1)+\frac{i}{t_1}\right) e^{i z_{\rm thr} t_1} \times \nonumber\\
\times [\delta(t_4)\delta(t_6)\delta''(t_5)-\delta'(t_4)\delta'(t_6)\delta(t_5)].
\label{mltf}
\end{align}
After multiplying~(\ref{mltf}) by the expansion of~(\ref{chiz}) and integrating with
respect to $\bmath t$, we derive the final Edgeworth-like $\FAP$
approximation~(\ref{FAPngauss}), along with the expressions for all coefficients $W_{ij}$
expressed as~(\ref{Wij}).

The quantities $q_{ij}$ in~(\ref{Wij}) are expressed via the cumulants ${\bmath l}_k$. In
practical computations, the latters can be estimated with $\mathcal O(\mu)$ accuracy as
follows. First, in the elements of $\bmath v$ all covariances are replaced with their
esimates~(\ref{covest}) and with analogous sample estimates for $\cov(y,y'')$ and $\var
y'$. Also, we must replace $Y$ by its estimation~(\ref{wavest}), as well as $Y'$ and $Y''$
should be replaced by the correponding derivatives of~(\ref{wavest}). This gives an
estimate $\tilde{\bmath v}$ that allows to compute arbitrary sample momenta of the
following kind:
\begin{equation}
L_{ij} = \langle \tilde v_i \tilde v_j \rangle, \; L_{ijk} = \langle \tilde v_i \tilde v_j \tilde v_k \rangle, \; L_{0000} = \langle \tilde v_0^4 \rangle - 3.
\label{kumest}
\end{equation}
By a convention, the indices in $v_i$ count from zero. For example:
\begin{align}
L_{11} &= \langle \tilde v_1^2 \rangle &\text{estimates} &\disp \frac{\partial y_{\rm nrm}}{\partial b}, \nonumber\\
L_{000} &= \langle \tilde v_0^3 \rangle &\text{estimates} &\expect y_{\rm nrm}^3 = \As y, \nonumber\\
L_{0000} &= \langle \tilde v_0^4 \rangle - 3 &\text{estimates} &\expect y_{\rm nrm}^4 - 3 = \Ex y, \nonumber\\
L_{011} &= \langle \tilde v_1^2 \tilde v_0 \rangle &\text{estimates} &\expect \left[ \left(\frac{\partial y_{\rm nrm}}{\partial b}\right)^2 y_{\rm nrm} \right].
\end{align}
All necessary quantities can be explicitly expressed via these $L_{\ldots}$. The most
simple expressions are:
\begin{align}
\det \mathbmss G &\simeq L_{11} L_{22} - L_{12}^2, \nonumber\\
q_{1,-3} &\simeq -L_{000}/3, \nonumber\\
q_{2,-6} &\simeq L_{000}^2/18.
\label{qij}
\end{align}
The formulae for $q_{02}$, $q_{1,-1}$, and $q_{2,-4}$ cannot be exposed here because of
their size, but they can be found in the MAPLE worksheet attached to the paper. They
contain various $L_{ij}$ and $L_{ijk}$, and a 4-index momentum $L_{0000}$ in $q_{2,-4}$.

\section{Uncertainty of the reconstructed density function}
\label{sec_recon}
Now it is convinient to use an equivalent form of the CWT~(\ref{wavdefkc}) and its
inverse~(\ref{wavinvkc}).

Let us assume that there is a `signal domain' $\mathcal S$ in the $(a,b)$ plane, in which
we put $Y_{\rm thr} = \widetilde Y$, while in all other parts of the plane $Y_{\rm
thr}=Y_0$. We can ssume that $\mathcal S$ is fixed. In actuality $\mathcal S$ is a random
outcome of the thresholding procedure that depends on the noise in $z(a,b)$, but in this
case there is an `average' signal domain, and the actual $\mathcal S$ may deviate by only a
small relative perturbation from it.

Let us substitute $Y_{\rm thr}(k,c)$ in~(\ref{wavinvkc}) and split the integration over
$\mathcal S$ and over its complement $\bar{\mathcal S}$:
\begin{align}
C_{\psi\gamma}\widetilde f(x) &= \iint\limits_{\bar{\mathcal S}} \Upsilon_0(\kappa,s) \gamma(\kappa x+s) d\kappa ds + \nonumber\\
&\quad + \iint\limits_{\mathcal S} \widetilde\Upsilon(\kappa,s) \gamma(\kappa x+s) d\kappa ds.
\label{fest}
\end{align}
The first integral in~(\ref{fest}) does not include any random noise, and it
represents a `predicted' part of the reconstructed p.d.f., i.e. the part for which the
model $Y_0$ appeared in good agreement with the sample. We denote this part as
$f_{\bar{\mathcal S}}(x)$. The random part is the second integral that
contains a noisy estimation $\widetilde\Upsilon$. Since it was defined as a sample
average~(\ref{wavest}), its mean
is $\expect\widetilde\Upsilon=\Upsilon$, and the formula for its covariance function can be
derived by generalizing~(\ref{wavvar}):
\begin{equation}
N \cov(\widetilde\Upsilon,\widetilde\Upsilon') = \int\limits_{-\infty}^{+\infty} f(x) \psi(\kappa x+s)\psi(k'x+c') dx - \Upsilon\Upsilon'.
\label{wavcovar}
\end{equation}
By using the equations~(\ref{wavdef}), (\ref{fest}), and~(\ref{wavcovar}), we can derive
the main statistical characteristics of the p.d.f. estimation $\widetilde f(x)$, namely its
mean and the variance:
\begin{align}
\expect \widetilde f(x) &= f_{\bar{\mathcal S}}(x)+ f_{\mathcal S}(x), \nonumber\\
N \disp \widetilde f(x) &= \int\limits_{-\infty}^{+\infty} R_{\mathcal S}^2(x,x') f(x') dx' - f_{\mathcal S}^2(x), \nonumber\\
f_{\mathcal S}(x) &= \int\limits_{-\infty}^{+\infty} R_{\mathcal S}(x,x') f(x') dx', \nonumber\\
R_{\mathcal S}(x,x') &= \frac{1}{C_{\psi\gamma}} \int\limits_{\mathcal S} \psi(\kappa x'+s) \gamma(\kappa x+s) d\kappa ds.
\label{festvar}
\end{align}

Both $\psi$ and $\gamma$ appearing in the definition of $R_{\mathcal S}$ are well-localized
functions, with the characteristic localization $\sim 1/\kappa=a$. Small $a$, or large
$\kappa$, dominate in the integral defining $R_{\mathcal S}(x,x')$, so this function must
be well localized with respect to the argument difference $x-x'$. Therefore, this kernel
basically performs a low-pass filtering on $f(x)$, and the filtering scale is about the
minimum value of $a$ present in the domain $\mathcal S$. This enables us to further
simplify~(\ref{festvar}) by employing the small-scale approximation, in which $f(x)$ can be
treated constant inside an integral:
\begin{align}
f_{\mathcal S}(x) &\simeq f(x) \int\limits_{-\infty}^{+\infty} R_{\mathcal S}(x,x') dx' = 0, \nonumber\\
N \disp \widetilde f(x) &\simeq f(x) \int\limits_{-\infty}^{+\infty} R_{\mathcal S}^2(x,x') dx'.
\end{align}

Now it becomes clear that the following aggregate noise characteristic gets a nice
representation:
\begin{equation}
N \int\limits_{-\infty}^{+\infty} \frac{\disp \widetilde f(x)}{f(x)} dx \simeq \int\limits_{-\infty}^{+\infty}\int\limits_{-\infty}^{+\infty} R_{\mathcal S}^2(x,x') dx dx'.
\label{fen}
\end{equation}
Its behaviour depends solely on the $L_2$ norm of the function $R_{\mathcal S}$, which
in turn depends on $\psi$ and $\gamma$. We must minimize $\|R_{\mathcal S}\|$ with respect
to $\gamma$ to obtain the least noisy p.d.f. reconstruction.

We can see that the optimality depends strongly on the domain $\mathcal S$
used in~(\ref{fen}). We need to specify $\mathcal S$ to proceed further, but
this depends on what kind of `optimality' we want from our reconstruction algorithm.

Thinking of this more deeply, the plain minimization of the cumulative noise~(\ref{fen})
does not appear very useful. In terms of our \emph{pattern} analysis, the distribution
$f(x)$ is viewed as a superposition of multiple \emph{patterns} with different scale and
position, and these patterns may overlap and interfere with each other. At least, there is
always a background large-scale structure that contrubutes in each value of $f(x)$ at any
given $x$, making it unclear what is the actual contribution from smaller-scale structures.
Instead of the `aggregate optimality', we would prefer to optimize individually all noise
contributions coming from each of the individual structures. This means the `local
optimality' that binds us to a more or less narrow domain in the $(a,b)$ plane.

Therefore, let us set $\mathcal S$ to a narrow box $\Delta\kappa \times \Delta s$ about a
point $(\kappa_0,s_0)$. This means to consider the contribution from just a single
elementary structure detected in the $(a,b)$ plane, and neglecting everything else
including the large-scale background. We have:
\begin{align}
R_{\mathcal S}(x,x') = \frac{\Delta\kappa \Delta s}{C_{\psi\gamma}} \psi(\kappa_0 x'+s_0) \gamma(\kappa_0 x+s_0), \nonumber\\
\|R_{\mathcal S}\| =  \frac{\Delta\kappa \Delta s}{\kappa_0 |C_{\psi\gamma}|}\, \|\psi\|\, \|\gamma\| \longmapsto \min_\gamma.
\end{align}
Note that $C_{\psi\gamma}$ also depends on $\gamma$, according to~(\ref{wavinv}). To
express it in a more convenient manner, let us define the auxiliary function $\eta$,
such that
\begin{equation}
\hat\eta(\omega) = \hat\psi(\omega)/|\omega|.
\end{equation}
Then we can write $C_{\psi\gamma}=(\eta,\gamma)$, meaning under $(*,*)$ the scalar product
in the Hilbert space. Thanks to the Parseval theorem, this scalar product can be
equivalently computed via the time-domain integration of $\eta\gamma$ or via the
frequency-domain integration of $\hat\eta^*\hat\gamma$. Using this notation, we must solve
\begin{equation}
\|R_{\mathcal S}\| \propto \frac{\|\gamma\|}{|(\eta,\gamma)|} \longmapsto \min_\gamma \iff
\frac{|(\eta,\gamma)|}{\|\gamma\|} \longmapsto \max_\gamma.
\end{equation}
Speaking in geometric wording, we must maximize the cosine of an `angle' between $\eta$ and
$\gamma$. This is obviously achieved if $\gamma$ is `parallel' to $\eta$, or
just put $\gamma=\eta$.

\section{Supplementary files: MAPLE worksheets}
\label{sec_maple}
The supplementary material contains two MAPLE worksheets:
\begin{enumerate}
\item {\sc edge.mw} - contains the full derivation of the FAP Edgeworth
decomposition~(\ref{FAPngauss}), following the scheme layed out in \ref{sec_edge};
\item {\sc qij.mw} - contains only the final expressions for $q_{ij}$ for the normality
test~(\ref{nrmtest}). They are expressed via the cumulants $l$ that are estimated by $L$
in~(\ref{kumest}).
\end{enumerate}

\bibliographystyle{model2-names}
\bibliography{wavelets}

\begin{thebibliography}{25}
\expandafter\ifx\csname natexlab\endcsname\relax\def\natexlab#1{#1}\fi
\expandafter\ifx\csname url\endcsname\relax
  \def\url#1{\texttt{#1}}\fi
\expandafter\ifx\csname urlprefix\endcsname\relax\def\urlprefix{URL }\fi
\providecommand{\eprint}[2][]{\url{#2}}
\providecommand{\bibinfo}[2]{#2}
\ifx\xfnm\relax \def\xfnm[#1]{\unskip,\space#1}\fi
\bibitem[{Abramovich et~al.(2000)Abramovich, Bailey and Sapatinas}]{ABS00}
\bibinfo{author}{Abramovich, F.}, \bibinfo{author}{Bailey, T.C.},
  \bibinfo{author}{Sapatinas, T.}, \bibinfo{year}{2000}.
\newblock \bibinfo{title}{Wavelet analysis and its statistical applications}.
\newblock \bibinfo{journal}{JRSS-D (The Statistician)} \bibinfo{volume}{49},
  \bibinfo{pages}{1--29}.
\bibitem[{Aza{\"\i}s and Delmas(2002)}]{AzaisDelmas02}
\bibinfo{author}{Aza{\"\i}s, J.M.}, \bibinfo{author}{Delmas, C.},
  \bibinfo{year}{2002}.
\newblock \bibinfo{title}{Asymptotic expansions for the distribution of the
  maximum of {G}aussian random fields}.
\newblock \bibinfo{journal}{Extremes} \bibinfo{volume}{5},
  \bibinfo{pages}{181--212}.
\bibitem[{Aza{\"\i}s and Wschebor(2009)}]{AzaisWschebor-levelsets}
\bibinfo{author}{Aza{\"\i}s, J.M.}, \bibinfo{author}{Wschebor, M.},
  \bibinfo{year}{2009}.
\newblock \bibinfo{title}{Level Sets and Extrema of Random Processes and
  Fields}.
\newblock \bibinfo{publisher}{Wiley}.
\bibitem[{Baluev(2008)}]{Baluev08a}
\bibinfo{author}{Baluev, R.V.}, \bibinfo{year}{2008}.
\newblock \bibinfo{title}{Assessing the statistical significance of periodogram
  peaks}.
\newblock \bibinfo{journal}{\mnras} \bibinfo{volume}{385},
  \bibinfo{pages}{1279--1285}.
\bibitem[{Baluev(2013)}]{Baluev13b}
\bibinfo{author}{Baluev, R.V.}, \bibinfo{year}{2013}.
\newblock \bibinfo{title}{Detecting non-sinusoidal periodicities in
  observational data: the von {M}ises periodogram for variable stars and
  exoplanetary transits}.
\newblock \bibinfo{journal}{\mnras} \bibinfo{volume}{431},
  \bibinfo{pages}{1167--1179}.
\bibitem[{Baluev(2017)}]{Baluev18b}
\bibinfo{author}{Baluev, R.V.}, \bibinfo{year}{2017}.
\newblock \bibinfo{title}{Fine-resolution wavelet analysis of exoplanetary
  distributions: hints of an overshooting iceline accumulation}.
\newblock \bibinfo{journal}{preprint},
  \bibinfo{pages}{arXiv.org:1712.06374}.
\bibitem[{Baluyev(2005)}]{Baluev05-51Peg}
\bibinfo{author}{Baluyev, R.V.}, \bibinfo{year}{2005}.
\newblock \bibinfo{title}{Statistics of masses and orbital parameters of
  extrasolar planets using continuous wavelet transforms}, in:
  \bibinfo{editor}{Arnold, L.}, \bibinfo{editor}{Bouchy, F.},
  \bibinfo{editor}{Moutou, C.} (Eds.), \bibinfo{booktitle}{Tenth anniversary of
  51 {P}eg -- b: {S}tatus of and prospects for hot {J}upiter studies},
  \bibinfo{publisher}{Frontier Group}, \bibinfo{address}{Paris}. pp.
  \bibinfo{pages}{103--110}.
\bibitem[{Brown et~al.(2016)}]{Gaia17}
\bibinfo{author}{Brown, A.G.A.}, et~al., \bibinfo{year}{2016}.
\newblock \bibinfo{title}{{G}aia {D}ata {R}elease 1. {S}ummary of the
  astrometric, photometric, and survey properties}.
\newblock \bibinfo{journal}{\aap} \bibinfo{volume}{595}, \bibinfo{pages}{A2}.
\bibitem[{Chereul et~al.(1998)Chereul, Cr\'{e}z\'{e} and
  Bienaym\'{e}}]{Chereul98}
\bibinfo{author}{Chereul, E.}, \bibinfo{author}{Cr\'{e}z\'{e}, M.},
  \bibinfo{author}{Bienaym\'{e}, O.}, \bibinfo{year}{1998}.
\newblock \bibinfo{title}{The distribution of nearby stars in phase space
  mapped by {H}ipparcos. {II}. {I}nhomogeneities among {A-F} type stars}.
\newblock \bibinfo{journal}{\aap} \bibinfo{volume}{340},
  \bibinfo{pages}{384--396}.
\bibitem[{Cumming(2010)}]{CummingStat}
\bibinfo{author}{Cumming, A.}, \bibinfo{year}{2010}.
\newblock \bibinfo{title}{Statistical distribution of exoplanets}, in:
  \bibinfo{editor}{Seager, S.} (Ed.), \bibinfo{booktitle}{Exoplanets}.
  \bibinfo{publisher}{University of Arizona Press}, \bibinfo{address}{Tucson}.
  chapter~\bibinfo{chapter}{9}, pp. \bibinfo{pages}{191--214}.
\bibitem[{Fadda et~al.(1998)Fadda, Slezak and Bijaoui}]{Fadda98}
\bibinfo{author}{Fadda, D.}, \bibinfo{author}{Slezak, E.},
  \bibinfo{author}{Bijaoui, A.}, \bibinfo{year}{1998}.
\newblock \bibinfo{title}{Density estimation with non–parametric methods}.
\newblock \bibinfo{journal}{\aaps} \bibinfo{volume}{127},
  \bibinfo{pages}{335--352}.
\bibitem[{Foster(1996)}]{Foster96c}
\bibinfo{author}{Foster, G.}, \bibinfo{year}{1996}.
\newblock \bibinfo{title}{Wavelets for period analysis of unevenly spaced time
  series}.
\newblock \bibinfo{journal}{\aj} \bibinfo{volume}{112},
  \bibinfo{pages}{1709--1729}.
\bibitem[{Grossman and Morlet(1984)}]{GrMorlet84}
\bibinfo{author}{Grossman, A.}, \bibinfo{author}{Morlet, J.},
  \bibinfo{year}{1984}.
\newblock \bibinfo{title}{Decomposition of {H}ardy functions into square
  integrable wavelets of constant shape}.
\newblock \bibinfo{journal}{SIAM J. Math. Anal.} \bibinfo{volume}{15},
  \bibinfo{pages}{723--736}.
\bibitem[{Hara et~al.(2017)Hara, Bou{\'e}, Laskar and Correia}]{Hara17}
\bibinfo{author}{Hara, N.C.}, \bibinfo{author}{Bou{\'e}, G.},
  \bibinfo{author}{Laskar, J.}, \bibinfo{author}{Correia, A.C.M.},
  \bibinfo{year}{2017}.
\newblock \bibinfo{title}{Radial velocity data analysis with compressed sensing
  techniques}.
\newblock \bibinfo{journal}{\mnras} \bibinfo{volume}{464},
  \bibinfo{pages}{1220--1246}.
\bibitem[{Horne and Baliunas(1986)}]{HorneBal86}
\bibinfo{author}{Horne, J.H.}, \bibinfo{author}{Baliunas, S.L.},
  \bibinfo{year}{1986}.
\newblock \bibinfo{title}{A prescription for period analysis of unevenly spaced
  time series}.
\newblock \bibinfo{journal}{\apj} \bibinfo{volume}{302},
  \bibinfo{pages}{757--763}.
\bibitem[{Jansen(2001)}]{Jansen-NRWT}
\bibinfo{author}{Jansen, M.}, \bibinfo{year}{2001}.
\newblock \bibinfo{title}{Noise Reduction by Wavelet Thresholding}.
\newblock \bibinfo{publisher}{Springer}.
\bibitem[{Liu et~al.(2015)Liu, Su and Wang}]{Liu15}
\bibinfo{author}{Liu, L.}, \bibinfo{author}{Su, X.}, \bibinfo{author}{Wang,
  G.}, \bibinfo{year}{2015}.
\newblock \bibinfo{title}{On inversion of continuous wavelet transform}.
\newblock \bibinfo{journal}{Open J. Stat.} \bibinfo{volume}{5},
  \bibinfo{pages}{714--720}.
\bibitem[{Martins(2010)}]{Martins10}
\bibinfo{author}{Martins, J.P.}, \bibinfo{year}{2010}.
\newblock \bibinfo{title}{Student t-statistic distribution for non-gaussian
  populations}, in: \bibinfo{booktitle}{Proc. ITI 2010, 32nd International
  Conference on Information Technology Interfaces}, \bibinfo{publisher}{IEEE}.
  pp. \bibinfo{pages}{563--568}.
\bibitem[{McEwen et~al.(2017)McEwen, Feeney, Peiris, Wiaux, Ringeval and
  Bouchet}]{McEwen17}
\bibinfo{author}{McEwen, J.D.}, \bibinfo{author}{Feeney, S.M.},
  \bibinfo{author}{Peiris, H.V.}, \bibinfo{author}{Wiaux, Y.},
  \bibinfo{author}{Ringeval, C.}, \bibinfo{author}{Bouchet, F.R.},
  \bibinfo{year}{2017}.
\newblock \bibinfo{title}{Wavelet-bayesian inference of cosmic strings embedded
  in the cosmic microwave background}.
\newblock \bibinfo{journal}{\mnras} \bibinfo{volume}{472},
  \bibinfo{pages}{4081--4098}.
\bibitem[{McEwen et~al.(2004)McEwen, Hobson, Lasenby and Mortlock}]{McEwen04}
\bibinfo{author}{McEwen, J.D.}, \bibinfo{author}{Hobson, M.P.},
  \bibinfo{author}{Lasenby, A.N.}, \bibinfo{author}{Mortlock, D.J.},
  \bibinfo{year}{2004}.
\newblock \bibinfo{title}{A high-significance detection of non-{G}aussianity in
  the {W}ilkinson {M}icrowave {A}nisotropy {P}robe 1-yr data using directional
  spherical wavelets}.
\newblock \bibinfo{journal}{\mnras} \bibinfo{volume}{359},
  \bibinfo{pages}{1583--1596}.
\bibitem[{Romeo et~al.(2003)Romeo, Horellou and Bergh}]{Romeo03}
\bibinfo{author}{Romeo, A.B.}, \bibinfo{author}{Horellou, C.},
  \bibinfo{author}{Bergh, J.}, \bibinfo{year}{2003}.
\newblock \bibinfo{title}{N-body simulations with two-orders-of-magnitude
  higher performance using wavelets}.
\newblock \bibinfo{journal}{\mnras} \bibinfo{volume}{342},
  \bibinfo{pages}{337--344}.
\bibitem[{Romeo et~al.(2004)Romeo, Horellou and Bergh}]{Romeo04}
\bibinfo{author}{Romeo, A.B.}, \bibinfo{author}{Horellou, C.},
  \bibinfo{author}{Bergh, J.}, \bibinfo{year}{2004}.
\newblock \bibinfo{title}{A wavelet add-on code for new-generation {N}-body
  simulations and data de-noising ({JOFILUREN})}.
\newblock \bibinfo{journal}{\mnras} \bibinfo{volume}{354},
  \bibinfo{pages}{1208--1222}.
\bibitem[{Schwarzenberg-Czerny(1998)}]{SchwCzerny98b}
\bibinfo{author}{Schwarzenberg-Czerny, A.}, \bibinfo{year}{1998}.
\newblock \bibinfo{title}{Period search in large datasets}.
\newblock \bibinfo{journal}{Baltic Astron.} \bibinfo{volume}{7},
  \bibinfo{pages}{43--69}.
\bibitem[{Skuljan et~al.(1999)Skuljan, Hearnshaw and Cottrell}]{Skuljan99}
\bibinfo{author}{Skuljan, J.}, \bibinfo{author}{Hearnshaw, J.B.},
  \bibinfo{author}{Cottrell, P.L.}, \bibinfo{year}{1999}.
\newblock \bibinfo{title}{Velocity distribution of stars in the solar
  neighbourhood}.
\newblock \bibinfo{journal}{\mnras} \bibinfo{volume}{308},
  \bibinfo{pages}{731--740}.
\bibitem[{Vityazev(2001)}]{Vit-wav}
\bibinfo{author}{Vityazev, V.V.}, \bibinfo{year}{2001}.
\newblock \bibinfo{title}{Wavelet analysis of time series (in Russian)}.
\newblock \bibinfo{publisher}{SPb Univ. Press}, \bibinfo{address}{Saint
  Petersburg}.

\end{thebibliography}







\end{document}